\pdfoutput=1
\documentclass{article}

\usepackage{arxiv}
\usepackage{color, colortbl}
\usepackage{comment}
\usepackage{graphicx}
\usepackage{subcaption}
\usepackage[labelfont=bf,textfont=bf]{caption}
\usepackage{balance}  
\usepackage{booktabs} 
\usepackage{float}
\usepackage{enumitem}
\usepackage{fancyvrb}
\usepackage{filecontents, pgffor}
\usepackage{hhline}    
\usepackage{multirow}
\usepackage{booktabs}
\usepackage{arydshln}
\usepackage{hhline}
\usepackage{diagbox}

\usepackage[noend,ruled, linesnumbered]{algorithm2e} 
\usepackage{mathtools}
\DeclarePairedDelimiter\ceil{\lceil}{\rceil}

\usepackage{pifont}
\newcommand{\cmark}{\ding{51}}%
\newcommand{\xmark}{\ding{55}}%
\usepackage{makecell} 
\usepackage{multirow} 
\usepackage{tabulary}
\usepackage{threeparttable}
\usepackage{array}
\usepackage{url}
\usepackage[T1]{fontenc}
\newcommand{\revision}[1]{{\color{black}{#1}}}

\usepackage{xspace} 

\author{
  Pingcheng Ruan \\
  National University of Singapore \\
  \texttt{ruanpc@comp.nus.edu.sg} \\
    \And
  Tien Tuan Anh Dinh \\
  Singapore University of Technology and Design \\
  \texttt{dinhtta@sutd.edu.sg} \\
    \And
  Dumitrel Loghin \\
  National University of Singapore \\
  \texttt{dumitrel@comp.nus.edu.sg} \\
    \And
  Meihui Zhang \\
  Beijing Institute of Technology \\
  \texttt{meihui\_zhang@bit.edu.cn} \\
    \And
  Gang Chen \\
  Zhejiang University \\
  \texttt{cg@zju.edu.cn} \\
    \And
  Qian Lin \\
  ByteDance \\
  \texttt{qian.lin@bytedance.com} 
    \And
  Beng Chin Ooi \\
  National University of Singapore \\
  \texttt{ooibc@comp.nus.edu.sg} 
}
\title{Blockchains vs. Distributed Databases: Dichotomy and Fusion}

\begin{document}
\maketitle

\begin{abstract}
Blockchain has come a long way --- a system that was initially proposed
specifically for cryptocurrencies is now being adapted and adopted as a
general-purpose transactional system.
As blockchain evolves into another data management system, the natural question
is how it compares against distributed database systems.
Existing works on this comparison focus on high-level properties, such as
security and throughput.
They stop short of showing how the underlying design choices contribute to the
overall differences.
Our work fills this important gap and provides a principled framework for
analyzing the emerging trend of blockchain-database fusion.


We perform a twin study of blockchains and distributed database systems as two
types of transactional systems.
We propose a taxonomy that illustrates the dichotomy across four dimensions,
namely replication, concurrency, storage, and sharding.
Within each dimension, we discuss how the design choices are driven by two
goals: security for blockchains, and performance for distributed databases. 
To expose the impact of different design choices on the overall performance, 
we conduct an in-depth performance analysis of two blockchains, namely Quorum
and Hyperledger Fabric, and two distributed databases, namely TiDB, and etcd.
\revision{Lastly, we propose a framework for back-of-the-envelope performance forecast 
of blockchain-database hybrids.}


\end{abstract}

\section{Introduction}
The very first blockchain system, that is Bitcoin~\cite{nakamoto2008bitcoin}, is a decentralized ledger for
recording cryptocurrency's transactions. The ledger consists of multiple blocks chained together with cryptographic
hash pointers, each block containing multiple transactions. This chain of blocks is distributed    
across a network of nodes, some of which behave in a Byzantine (or malicious) manner~\cite{Lamport_BFT}. The network runs a
consensus protocol, namely \textit{proof-of-work} (PoW), to keep the ledger consistent among the nodes.  

Bitcoin is the first digital currency (or cryptocurrency) system that
operates in a Byzantine \cite{Lamport_BFT} peer-to-peer (P2P) environment, without relying on a common trusted
third party.
But it can execute only simple transactions that move some coins from one address (or user) to another.
Recent blockchains such as Ethereum~\cite{wood2014ethereum} and Hyperledger
Fabric~\cite{androulaki2018hyperledger} support general-purpose transactions. The key enabler is the {\em
smart contract} --- a user-defined computation executed by all nodes in the blockchain. 
With smart contracts, blockchains can execute transactional workloads which have so far been handled almost
exclusively by 
databases.  In other words, blockchains have evolved into transactional management systems, and therefore are
comparable to distributed databases. Their advantages over the latter include data transparency and security
against Byzantine failures. 
In fact, many companies and government agencies are exploring blockchains to replace, or to complement, their
enterprise-grade databases~\cite{mougayar2016business,morabito2017business,crosby2016blockchain}.

The parallel between blockchains and distributed databases has not gone
unnoticed. Existing works show that there are little similarities between the
two. Blockchains are suitable when the applications are running in untrusted,
hostile environments, whereas databases are suitable when performance is more
important than security~\cite{crosby2016blockchain,
wust2018you,chowdhury2018blockchain,yaga2018blockchain}. Their distinction is
further compounded by the significant gap in
performance~\cite{dinh2017blockbench}, for instance Bitcoin processes around
$10$ transactions per second~\cite{bitcoin_tps} while etcd --- a
state-of-the-art distributed NoSQL database --- processes over $50,000$
operations per second~\cite{etcd_perf}.

On the other hand, we notice the trend of design fusion between databases and
blockchains. Design principles and techniques that are traditionally used by
databases are being adopted by blockchains. For example, concurrency control
techniques attributed to databases are used to increase the performance of
blockchains~\cite{sharma2019blurring, ruan2020transactional,
dickerson2017adding}. Moreover, sharding has been used to scale out permissioned
blockchains~\cite{dang2018towards}. At the same time, the security features of
blockchains are used in hybrid blockchain-database systems to provide verifiable
data~\cite{el2019blockchaindb, veritas, peng2020falcondb}.

One limitation of the existing works that compare blockchains and databases is
that they only focus on application-level, observable and measurable properties,
such as throughput and security. In particular, they show how the two types of
systems differ without identifying the root cause. For example,
BLOCKBENCH~\cite{dinh2017blockbench} compares three private blockchains, namely
Hyperledger Fabric, Ethereum and Parity, with H-Store under two popular data
processing workloads. The authors expose a large gap in performance, but provide
no further analysis of that gap. As a consequence, the reported difference does
not generalize to workloads other than the two used in the experiments. For
instance, under high contention workloads, the performance difference may shrink
drastically. 

To overcome these limitations, we aim to provide a comprehensive dichotomy of
blockchains and databases. Our approach is to position them within the same
design space --- that is, the design space of general transactional systems. We
propose a taxonomy consisting of four design dimensions and discuss how the two
types of systems make different design choices in each dimension. The first
dimension is replication, which determines what data is replicated to what
nodes, and the mechanism needed to keep the replicas consistent. The second is
concurrency, which determines the tradeoffs between performance and correctness
when executing concurrent transactions. The third is storage, which determines
the data models and access methods. The final dimension is sharding, which
determines how data is partitioned, and the mechanism for atomicity of
cross-shard transactions.

\revision{The four dimensions in our taxonomy capture the fundamental
similarities betwen blockchains and databases. In addition, their impact on the
overall performance can be measured, therefore these dimensions form a framework
for fine-grained, quantitiative comparison between these systems. We demonstrate
how our taxonomy is useful in practice by applying it to compare the performance
of recent hybrid database-blockchain
systems~\cite{BlockchainMeetsDatabase,peng2020falcondb,veritas,el2019blockchaindb,mcconaghy2016bigchaindb,schuhknecht2019chainifydb}.}

In summary, we make the following contributions in this paper:
\begin{itemize}
  \item We compare blockchains and distributed databases as two different types
  of distributed, transactional systems. We propose a new taxonomy that
  characterizes both types of systems and their hybrids along four
  design dimensions: replication, concurrency, storage, and sharding.

  \item We conduct a comprehensive performance study of four popular systems,
  including two permissioned blockchains, namely Fabric~\cite{web:fabric} and
  Quorum~\cite{web:quorum}, and two database systems, namely
  TiDB~\cite{web:tidb} and etcd~\cite{web:etcd}. The results demonstrate the
  impact of different design choices on performance.
   
  \item We use our taxonomy to analyze the security and performance of emerging
  hybrid blockchain-database systems. We propose a framework that explains their
  performance differences and estimates the performance of future hybird
  systems.
\end{itemize}

Section~\ref{sec:background} provides relevant background, followed by a qualitative comparisons on the above
four dimensions in Section~\ref{sec:taxonomy}. Section~\ref{sec:setup} and Section~\ref{sec:result} discuss the experimental setup and
results, respectively. Section~\ref{sec:related} reviews related works before Section~\ref{sec:conclusions} concludes. 

\section{Background}
\label{sec:background}

In this section, we discuss relevant background on blockchains and distributed databases.
Figure~\ref{fig:spectrum} shows a high-level comparison.

\subsection{Blockchain}

From a data structure perspective, a blockchain is a list of blocks linked by
cryptographic hash pointers. These blocks contain cryptocurrency
transactions~\cite{nakamoto2008bitcoin}. By this definition, the blockchain is a
tamper-evident ledger for recording transactions. With smart contracts,
transactions are in the form of contract deployment and invocation.
From a systems perspective, a blockchain is a distributed system consisting of
multiple nodes, some of which are malicious. These nodes maintain a consistent
ledger by using a Byzantine fault-tolerant (BFT) consensus protocol, such as PoW
or PBFT~\cite{castro1999practical}.

\begin{figure*}[tp]
  \centering
  \includegraphics[width=0.75\textwidth]{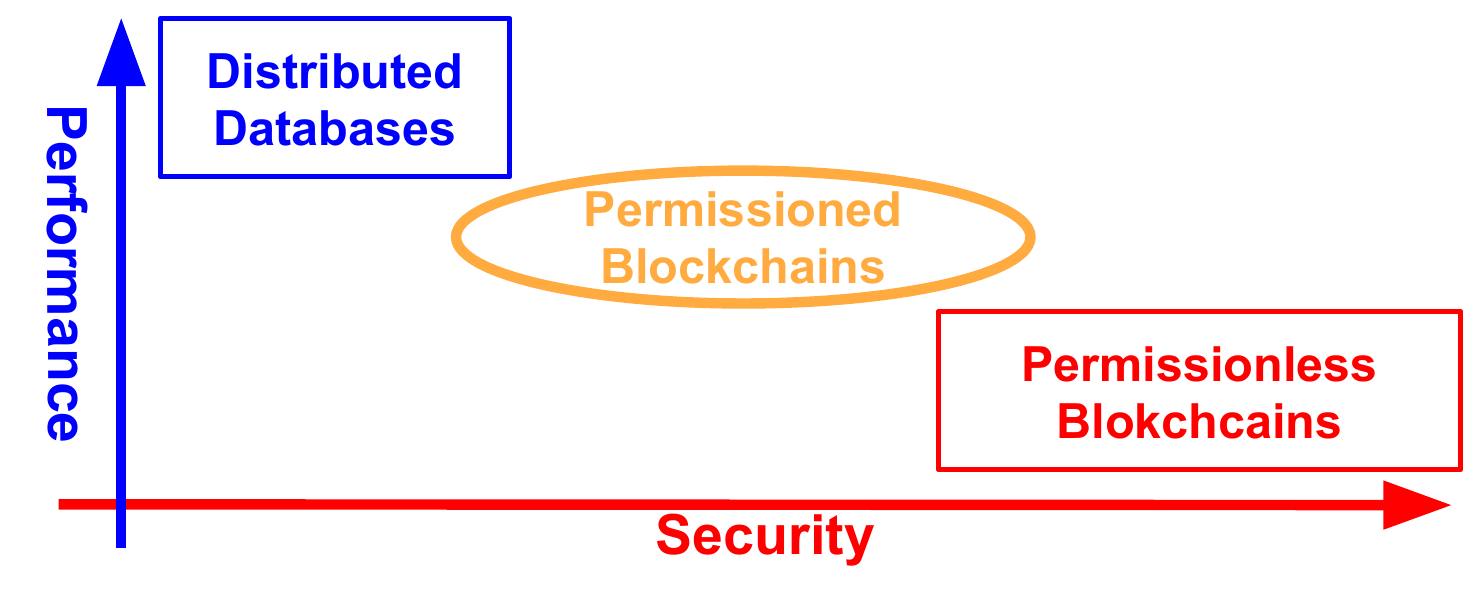}
  \caption{Blockchains vs. distributed databases in the security-performance coordinate.}
  \label{fig:spectrum}
 \end{figure*}

In earlier designs, a blockchain transaction is restricted to cryptocurrency and
the states are modeled as Unspent Transaction Outputs (UTXO).  For example,
Bitcoin~\cite{nakamoto2008bitcoin} and other similar altcoins use the UTXO
model.  Starting with Ethereum~\cite{web:ethereum}, blockchains support
\textit{smart contracts} which allow users to encode and execute arbitrary
Turing-complete computations on the ledger. The ledger states are modelled as
accounts instead of UTXO.
Other systems supporting smart contracts include Quorum, Parity and Hyperledger
Fabric~\cite{androulaki2018hyperledger}.  In these systems, a transaction on the
ledger takes the form of a contract invocation.
Transactions sequentially modify the system state based on their order in the
ledger, determined by the consensus protocol. A read-only transaction can be
carried out by any node, without undergoing the consensus and being included in
the ledger. We only consider blockchains that support smart contracts in this
paper, because earlier blockchains (without smart contracts) cannot support
database transaction workloads and thus cannot be compared with distributed
databases.


\textbf{Permissionless vs. Permissioned.} Blockchains can be broadly divided
into two categories: permissionless (or public), and permissioned (or private).
In the former, for example in Bitcoin and Ethereum, any node and user can join
the system in a pseudonymous manner. In the latter, for example in Fabric and
Quorum, the node and user must be authorized to join the system. With strong
membership control and action regulation, permissioned blockchains are more
suitable for enterprise applications and are particularly used in the financial
sector.
Figure~\ref{fig:spectrum} shows the security-performance tradeoffs in
blockchains. It highlights how permissionless blockchains can achieve stronger
security because they make no identity assumption. In contrast, permissioned
blockchains have weaker security because of the identity assumption, but can
achieve higher performance because they can employ consensus protocols with
higher efficiency. A more detailed discussion of permissionless versus
permissioned blockchain designs can be found
in~\cite{dinh2018untangling,dinh2017blockbench}.

\subsection{Distributed Databases}

Unlike blockchains, database systems have been around for decades.  Relational
databases, which support easy-to-use SQL language and intuitive ACID transaction
semantics, remained mainstream throughout the years. The recent demand of big
data processing and the fact that Moore's law is reaching its limit are major
factors behind the trend of scale-out database designs. Nowadays, both data and
computation are distributed over multiple nodes in order to achieve high
availability and scalability. Principles and techniques in designing and scaling
distributed databases are described in detail in~\cite{ozsu2011principles}.
Basically, there are two distinctive movements, namely NoSQL and NewSQL, under
this new distributed design direction.

\textbf{NoSQL vs. NewSQL.}
For scalability, many distributed databases abandon the complex relational model
and the strong ACID semantics.
These systems are referred to as {\em NoSQL}. They support more flexible data
models and weaker consistency.
In the sense of the CAP theorem~\cite{gilbert2012perspectives}, these NoSQL
systems compromise consistency for the sake of availability.
A variety of their supported data models include key-value store (e.g,
Redis~\cite{carlson2013redis}, etcd~\cite{web:etcd}), document store (e.g,
CouchDB~\cite{anderson2010couchdb}), graph store (e.g,
Neo4J~\cite{vukotic2014neo4j}), column-oriented (e.g,
Cassandra~\cite{lakshman2010cassandra}) and so on.
The most lenient consistency model is eventual consistency which makes no
guarantees about the order of read and write operations.
Between eventual and strong consistency, researchers explore a variety of other
abstractions, such as sequential, causal, and PRAM consistency. They standardize
on the allowable operation behavior for the ease of reasoning.
Most NoSQL databases offer configurable options, where users can trade off
between performance and consistency.

The surge of NoSQL systems, however, does not obscure the cost in usability and
the increase in application complexity.  A new class of distributed database
systems, called {\em NewSQL}, aim to restore the relational model and ACID
semantics without sacrificing much scalability.
NewSQL has drawn attention since Google introduced
Spanner~\cite{corbett2013spanner}, the first NewSQL system. It was followed by a
few database vendors, such as CockroachDB~\cite{web:cockroach} and
TiDB~\cite{web:tidb}. In this paper, we consider both NoSQL and NewSQL systems.

\begin{figure*}[tp]
    \centering
	\begin{subfigure}{0.45\textwidth}
	    \includegraphics[width=0.9\textwidth]{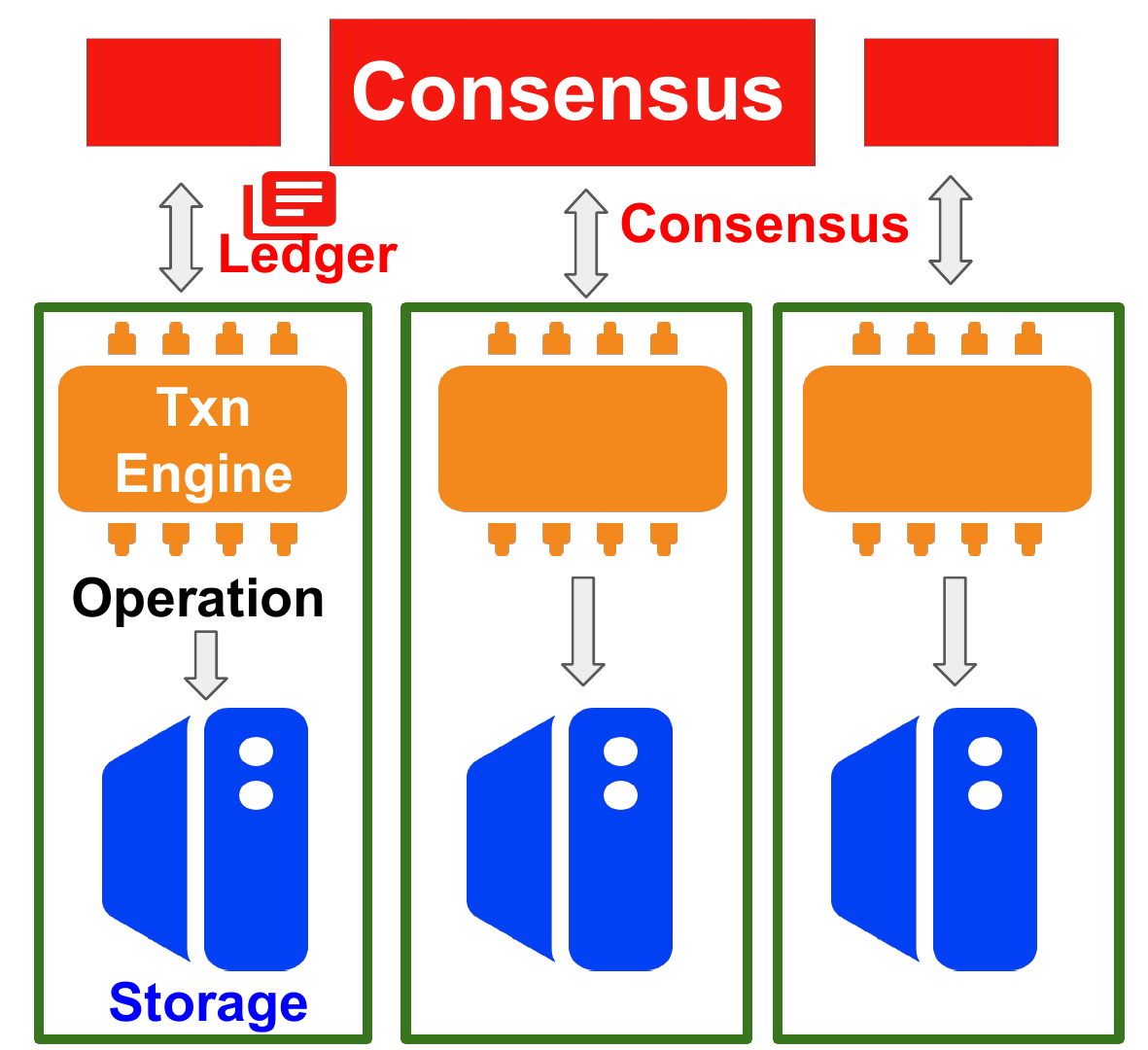}
	    \caption{}
	\end{subfigure}%
	\begin{subfigure}{0.45\textwidth}
	    \includegraphics[width=0.9\textwidth]{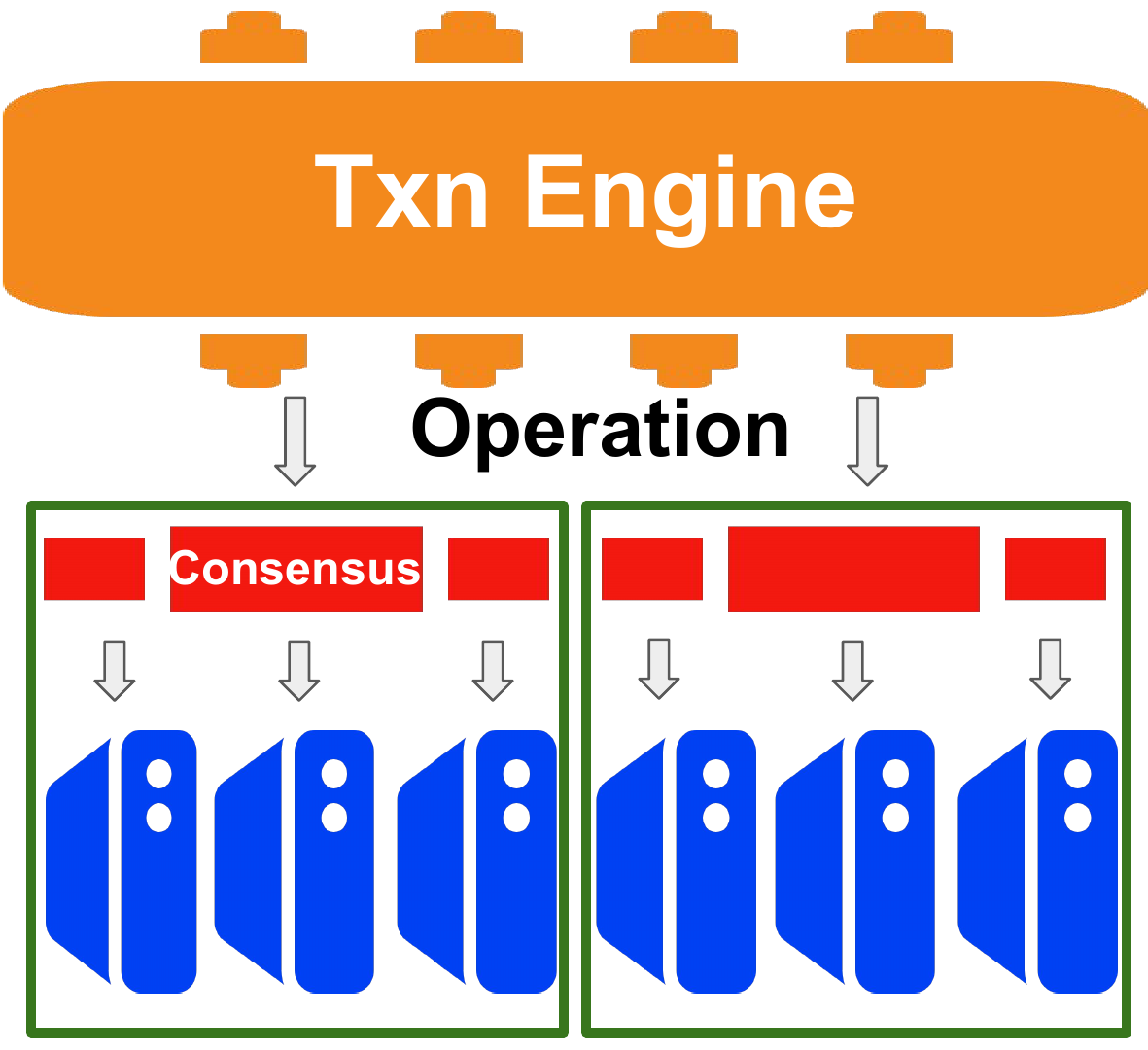}
	    \caption{}
	\end{subfigure}%
    \caption{(a) Blockchains first reach consensus on the transaction history, then commit their effects into the storage.
    (b) Distributed databases replicate at the storage layer.}
    \label{fig:arch}
\end{figure*}


\begin{table*}[tp]
\centering
\caption{Design choices in blockchains and distributed databases}
\label{tab:taxonomy}
\resizebox{\textwidth}{!}{%
\begin{tabular}{l|ll}
\toprule
& \textbf{Blockchains} & \textbf{Distributed Databases} \\ 
\hline
\textbf{Replication} & \begin{tabular}[c]{@{}l@{}}Transaction replication \\
Byzantine Fault Tolerant consensus: PBFT, PoW, etc.\end{tabular} &
\begin{tabular}[c]{@{}l@{}}State replication \\ 
Crash Fault Tolerant consensus: Raft, Paxos, etc.\end{tabular} \\ 
\hline \textbf{Concurrency} & Serial execution & Concurrent execution \\ 
\hline \textbf{Storage} & \begin{tabular}[c]{@{}l@{}}Append-only ledger
abstraction \\ 
Authenticated data structure: Merkle Tree, etc.\end{tabular} &
\begin{tabular}[c]{@{}l@{}}Direct access without historical query \\
Hardware-conscious index: PSL, FAST, etc.\end{tabular} \\ 
\hline \textbf{Sharding} & \begin{tabular}[c]{@{}l@{}}Node-aware shard
formation \\
2PC and BFT-based replication\end{tabular} & \begin{tabular}[c]{@{}l@{}}Workload-aware shard formation \\ 
2PC with centralized coordinator\end{tabular} \\
\bottomrule
\end{tabular}
}
\end{table*}

\section{Taxonomy}
\label{sec:taxonomy}

Table~\ref{tab:taxonomy} compares the design choices of distributed databases and blockchains under each dimension in our taxonomy. 

\subsection{Replication}
Replication is the technique of storing copies of the data on multiple nodes called replicas. The key
challenge in such a system is to ensure consistency under failures.
In this section, we characterize blockchains and distributed databases by what they replicate, how they keep
the replicas consistent, and their failure models.

\subsubsection{Replication model}
The units of replication can be transactions or the read/write operations.
Figure~\ref{fig:arch} shows that blockchains replicate an ordered log of transactions (or ledger).  
Distributed databases replicate the ordered log of read and write operations on top
of the storage. The nodes in the database are oblivious to the transaction logic because they see only one
operation at a time. One consequence of this model is that the transaction manager which coordinates the
execution of a transaction must be trusted. In contrast, a blockchain does not have such trusted entity,
therefore it replicates the entire transaction so that its execution can be replayed by each participant node.

By replicating transactions, the ledger contain application-level information, such as transaction context,
client signature, execution timestamp, etc.,  making it easy to perform transaction verification.  Due to this
verifiability, blockchains are often used as a data and computing platform for mutually distrusting parties.
On the other hand, replicating storage operations means there can be more concurrency, because
operations can be replicated in different order but with the same effect on the storage.  


\subsubsection{\revision{Replication approach}}
There are two main approaches to maintain consistency among replicas. The first
is primary-backup, which dedicates a replica as the primary which synchronizes
its states with backup replicas. This is adopted by many databases.
For example, Replex~\cite{replex} that uses chain replication.
Cassandra~\cite{lakshman2010cassandra} uses the client as the primary to
synchronize the replicas.

The second approach is state-machine replication, which essentially maintains an
ordered log of operations/transactions on each replica. Each replica starts at
the same initial state, then applies the operations/transactions in the log in
the same order. Many systems use {\em consensus protocols}, such as
Paxos~\cite{lamport2001paxos}, Raft~\cite{raft}, and
PBFT~\cite{castro1999practical}, for the replicas to agree on the ordered log.
Examples include Quorum~\cite{web:quorum}, TiDB~\cite{web:tidb},
Spanner~\cite{corbett2013spanner}.
Compared with the primary-backup, consensus protocols achieve the automatic
primary failover, by introducing the view change. That is, when the progress
halts, replicas may jointly agree to enter into a new epoch/view where a new
primary is elected.
Apart from the consensus, other systems rely on external services that provide a
distributed {\em shared log} abstraction, such as Kafka~\cite{web:kafka} and
Corfu~\cite{corfu}. Operations/transactions are appended to the log, and the
replicas, as clients of the log, apply them independently.
Examples of shared log systems include Fabric~\cite{web:fabric},
Hyder~\cite{hyder}, and Tango~\cite{tango}.

Primary-backup protocols are simpler, and can perform better than state-machine
replication, when the states are small and there are no failures. For example,
the chain replication protocol can spread the network cost more evenly among the
replicas than a consensus protocol, and it achieves better read
performance~\cite{replex}. Systems based on shared log are expected to perform
better than the ones based on consensus when there are no failures.  This is
because shared log decouples ordering from state replication, therefore it can
be optimized to have high throughput. Furthermore, while the throughput of a
consensus protocol decreases with more replicas, the throughput of a shared log
system is expected to remain constant until the number of log consumers exceeds
the capacity of the log producers~\cite{corfu}. 
 


\subsubsection{Failure model}
\label{sec:taxonomy:replication:failure}
Replication protocols are complex because they need to maintain consistency
under failure. Under the crash failure model, in which nodes only fail by
crashing, the protocols need to tolerate hardware and software failures. Under
the Byzantine failure model, in which nodes fail arbitrarily, the protocols need
to tolerate any software and hardware failures, as well as any malicious
behavior. This model is suitable for achieving security, since it considers
attacks that fully compromise some nodes in the system.

An orthogonal dimension to the node failure model is the network assumption. The
network is {\em synchronous} when the network delay is bounded and known. It is
{\em asynchronous} when the network delay is unbounded. Protocols that tolerate
crash failures, or CFT, require $f+1$ replicas to tolerate $f$ failures under
the synchronous network model~\cite{budhiraja1993primary}, and $2f+1$ under the
asynchronous network model~\cite{raft,lamport2001paxos}. Protocols that tolerate
Byzantine failure, or BFT, require $2f+1$ and $3f+1$ replicas to tolerate $f$
failures under the synchronous and asynhronous network models,
respectively~\cite{castro1999practical, yin2019hotstuff,buchman2016tendermint}.


Databases assume the crash failure model, since they are considered internal
systems which are not subject to security attacks. For example,
Spanner~\cite{web:spanner} uses Paxos~\cite{lamport2005generalized}, a CFT
protocol.  Permissioned blockchains support both failure models. For example,
Quorum provides implementations for both Raft~\cite{raft}, a CFT protocol, and
IBFT, a BFT protocol. These systems allow applications to make different
tradeoffs between security and performance.
Public blockchains, on the other hand, ubiquitously adopt BFT protocols because they admit
any nodes to the system. In particular, PoW protocols are often used because
they address one fundamental problem in the public settings: a node can have
many identities. In PoW, a node's probability of solving a computational puzzle,
thereby reaching consensus and gaining rewards, is proportional to its physical
resources which are difficult to forge.



In an asynchronous network and under failures, the FLP
theorem~\cite{fischer1982impossibility} rules out any deterministic consensus
protocol that can achieve both safety and liveness. Public blockchains choose
liveness over safety, meaning that the systems remain available under network
partitions, but these partitions may be in disagreement which takes the form of
forks.
\revision{Here, availability refers to the system's behavior, that is, new
blocks are appended to the ledger. Individual transactions may be
censored and excluded from the blockchain.}



PoW protocols have low throughputs mainly due to the resource
requirements~\cite{dinh2017blockbench}.
CFT protocols have better performance than BFT protocols, because the former
incur $O(N)$ network cost, whereas the latter incur $O(N^2)$, where $N$ is the
number of nodes. As a result, BFT protocols do not scale to a large number of
nodes, and their performance is more sensitive to network conditions at scale.
Specifically, when $N$ is large, BFT protocols are more likely to enter view
change --- an expensive phase of the protocol for replacing the current leader.



\subsection{Concurrency}

Concurrency refers to the extent to which transactions are executed at the same
time. There are two choices: transactions are executed either serially (or
sequentially), or concurrently. Most blockchains support only serial execution,
while distributed databases employ sophisticated concurrency control mechanisms
to extract as much concurrency as possible.

There are two reason behind blockchains' lack of support for concurrency. First,
serial execution may not affect the overall performance because transaction
execution is often not the bottleneck~\cite{dinh2017blockbench}.  \revision{ For
example, in Bitcoin, the consensus protocol may take several minutes to
complete, which is the block interval required by the protocol, whereas the
transaction execution, which invokes the Bitcoin script to validate a
cryptocurreny flow, can be done in milliseconds.} Second, serial execution means
the behavior of smart contracts is deterministic when the transaction execution
is replicated over many nodes. The benefit of determinism is that it is easy to
reason about the states of the ledger.

Unlike blockchains, concurrency remains a major research topic in databases, as
it is the main source of performance improvement. The challenge in extracting
concurrency is to ensure the correctness of the concurrent execution. In fact,
there is a wide range of isolation
levels~\cite{computer1986american,bailis2013highly} which make different
tradeoffs between correctness and performance. Most production-grade databases
today offer more than one isolation level.

We observe that recent blockchains are adopting some simple concurrency
techniques often found in databases. In Hyperledger Fabric, for example,
transactions are simulated (executed) in parallel against the ledger states
before being sent for ordering. During the later commit phase, the system uses a
simple optimistic concurrency control to achieve serializability which aborts
transactions whose simulated states are stale. More established techniques to
reduce abort have also been proposed~\cite{sharma2019blurring,
ruan2020transactional}.

\subsection{Storage}

\subsubsection{Storage model}
Storage can be built upon the latest states only, amenable for mutation, or upon
all historical information, amenable for appending.
The storage in distributed databases only exposes direct access to up-to-date
records. In databases without explicit provenance support, historical data is
maintained in limited forms, for example as write-ahead logs.
We note that such logs are used primarily for failure recovery, and they are
periodically pruned.
Blockchains, besides the state storage, additionally expose an append-only
ledger abstraction.
The ledger, a chain of blocks, records historical transactions and the changes
made to the global states.
We note that such a ledger is hash protected to conserve historical integrity.
Some blockchains allow applications to access only the latest states, for
example, Hyperledger Fabric v0.6.
Recently, novel blockchain-tailored storage systems have been proposed to enable
access to any historical states during smart contract
execution~\cite{lineagechain}.

\subsubsection{Index}
Indexes play an instrumental role on the state storage to facilitate data
access.
Apart from the performance consideration, some security-oriented systems
additionally rely on the index to compute a digest, which uniquely identifies
the state contents.
Distributed databases are more concerned by performance, i.e., any small
optimization on the index can translate to a significant improvement in
performance. Modern indexes are designed to be hardware-conscious in order to
extract the most efficiency from the hardware. For example, in-memory databases
abandon the disk-friendly B-tree structure for other structures such as
FAST~\cite{kim2010fast} and PSL~\cite{xie2017parallelizing} which are designed
for better cache utilization and multi-core parallelism.

To compute the content-unique digest, blockchains employ an authenticated data
structure, such as the Merkle tree index, to provide integrity protection on top
of the state storage.
For example, Ethereum uses a prefix trie, named Merkle Patricia Trie
(MPT)~\cite{web:mpt}. In MPT, the states are stored in the leaves. The states
with a common key prefix are organized under the same branch. Each node is
associated with the cryptographic hash of its content in the storage engine,
such that the root hash represents the complete global states. The access path
serves as the integrity proof for the retrieved value. Older versions of
Hyperledger Fabric use a Merkle Bucket Tree (MBT) in which the size of the tree
is fixed.
Unlike the ledger abstraction which is ubiquitous in blockchains, we note that
not all blockchains adopt the authenticated data structure for the state
organization. For example, Hyperledger Fabric abandons this design from version
1 onwards.	

\subsection{Sharding}

Sharding is a common technique in distributed databases for achieving
scalability, in which data is partitioned into multiple shards. Although it has
been studied extensively in databases, sharding has only recently been
introduced to blockchains \revision{to harness concurrency across shards}. In
this section, we discuss two key challenges in any sharded system, that are (i)
how to form a shard, and (ii) how to ensure atomicity for cross-shard
transactions.

\subsubsection{Shard formation}
A shard formation protocol determines which nodes and data go to which shard.
The security of blockchains depends on the assumption that the number of
failures is below a certain threshold. The shard formation protocol must,
therefore, ensure that the assumption holds for every shard.
In particular, the shard size must be large enough so that the fraction of
Byzantine nodes is small.
Furthermore, the attacker must not be able to influence the shard assignment,
otherwise, it could reserve enough resources for one shard to break the security
assumption. State-of-the-art sharded blockchains have different approaches. For
example, Elastico uses PoW for shard formation~\cite{luu2016secure}, while
\revision{ the recent version of Ethereum uses Proof-of-Stake to select
validators for each shard~\cite{web:eth2}.}
OmniLedger~\cite{kokoris2018omniledger} employs a complex cryptographic
protocol, while AHL~\cite{dang2018towards} uses trusted hardware.
\revision{These protocols are secured against Sybil attacks, and executed
regularly, in the form of shard reconfiguration, to guard against adaptive
attackers.}

The goal of sharding in distributed databases is scalability. As such, the
systems aim to assign data to shards in a way that optimizes the performance of
certain workloads. In practice, they offer a variety of partitioning schemes,
for example, hash partitioning and range partitioning, so that users can select
the most suitable for their workloads. Some systems, for instance,
Cassandra\cite{lakshman2010cassandra}, even allow users to specify workload
distributions so that data can be partitioned in a locality-aware manner. Unlike
blockchains, shard reconfiguration is not necessary for databases, unless when
there are significant changes in the workload distribution.

\subsubsection{Atomicity}
Sharding introduces the problem of transaction atomicity when a transaction
touched data in multiple shards.
Atomicity requires a cross-shard transaction to either commit or abort in all
shards. In databases, this problem is addressed by the two-phase commit (2PC)
protocol. This protocol requires a dedicated transaction coordinator that must
be trusted, but may fail and leave the transaction blocked forever. A recent
work proposed Parallel Commit to reduce the commit duration to a single round
trip~\cite{taft2020cockroachdb}.

Sharded blockchains face additional challenges in ensuring atomicity because the
coordinator cannot be trusted under the Byzantine failure model.
To overcome this, \revision{Eth2 introduces a separate chain running Casper
consensus~\cite{buterin2017casper}, called Beacon Chain, that coordinates
cross-shard transactions~\cite{web:eth2}}.
Similarly, \cite{dang2018towards,herlihy2019cross} propose to implement the 2PC
coordinator as a state machine in a shard that runs a BFT protocol. The BFT
protocol ensures that the shard is less vulnerable to attacks and does not
become a point of failure.
Any cross-shard transaction must involve this 2PC BFT replicated state machine
to ensure atomicity.
The consensus liveness guarantees the high availability of the coordinator,
therefore mitigating the blocking problem.
But the Byzantine setup in blockchains imposes considerable overhead to the 2PC
process.

\begin{table*}[tp]
	\centering
	\caption{\revision{System comparison based on our taxonomy with the benchmarked ones and their versions highlighted. 
	For each hybrid system, we mark their security-oriented designs with red and performance-oriented designs with blue.
	}}
	\label{tab:systems}
	
	\definecolor{highlighted}{rgb}{0.7, 0.75, 0.71}

	\newcommand{\security}[1]{{\color{red}{#1}}}
	\newcommand{\performance}[1]{{\color{blue}{#1}}}

	\resizebox{\textwidth}{!}{%
	\begin{tabular}{cclcccccc} 
		\toprule
		\multirow{2}{*}{\textbf{Category}} & \multirow{2}{*}{\textbf{System}} & \multicolumn{3}{c}{\textbf{Replication} }                                                                                                                                                  & \multirow{2}{*}{\textbf{Concurrency} } & \multicolumn{2}{c}{\textbf{Storage} } & \multirow{2}{*}{\begin{tabular}[c]{@{}c@{}}\textbf{Sharding }\\\textbf{ Support (2PC)} \end{tabular}}  \\
								  &                          & \begin{tabular}[c]{@{}l@{}}Replication \\ Model \end{tabular} & \begin{tabular}[c]{@{}c@{}}Replication \\Approach \end{tabular} & \begin{tabular}[c]{@{}c@{}}Failure Model\\(Consensus Protocol) \end{tabular} &                                        & Ledger Abstraction & Index(Storage Engine)                 &                                                                                                  \\

		\midrule
		\multirow{2}{*}{\makecell{Permissionless \\ Blockchains}} & Ethereum~\cite{web:ethereum} & Txn-based & Consensus & BFT(PoW) & Serial & \cmark & LSM Tree(LevelDB)+MPT & \xmark(\xmark)\\

		\cdashline{2-9} \\
		& Eth2~\cite{web:eth2} & Txn-based & Consensus & BFT(PoS + Casper) & Serial (in each shard) & \cmark & LSM Tree(LevelDB)+MPT & \cmark(\xmark)\\

		\midrule

		\multirow{5}{*}{ \makecell{Permissioned \\ Blockchains}} & \cellcolor{highlighted}Quorum v2.2~\cite{web:quorum} &  \cellcolor{highlighted}Txn-based &  \cellcolor{highlighted}Consensus &  \cellcolor{highlighted}Raft(CFT)/IBFT(BFT) &  \cellcolor{highlighted}Serial &  \cellcolor{highlighted}\cmark & \cellcolor{highlighted} LSM Tree(LevelDB)+MPT &  \cellcolor{highlighted}\xmark(\xmark) \\

		\cdashline{2-9} \\

		&\cellcolor{highlighted} Fabric v2.2~\cite{web:fabric} &\cellcolor{highlighted} Txn-based &\cellcolor{highlighted} Shared log &\cellcolor{highlighted} CFT(\textit{Orderer} with Raft) &\cellcolor{highlighted} \makecell{Concurrent Execution \\ Serial Commit}&\cellcolor{highlighted} \cmark &\cellcolor{highlighted} LSM Tree(LevelDB) &\cellcolor{highlighted} \xmark(\xmark) \\

		\cdashline{2-9} \\

		& Fabric v0.6~\cite{web:fabric06} & Txn-basformatsed & Consensus & BFT(PBFT) & Serial & \cmark & \makecell{LSM Tree(RocksDB) \\ + MBT} & \xmark(\xmark)
		\\

		\cdashline{2-9} \\
		& EOS~\cite{web:eos} & Txn-based & Consensus & BFT(DPoS) & Serial & \cmark & B-tree(MongoDB) &  \xmark(\xmark) \\

		\cdashline{2-9} \\

		& \multirow{2}{*}{FISCO BCOS~\cite{web:fisco}} & \multirow{2}{*}{Txn-based} &  Consensus & CFT(Raft) & \multirow{2}{*}{Serial} & \multirow{2}{*}{\cmark} & \multirow{2}{*}{\makecell{LSM Tree(LevelDB) \\+MPT}} & \multirow{2}{*}{\xmark(\xmark)} \\
		& &  & & BFT(PBFT) & &  &  &  \\

		\midrule
		\multirow{3}{*}{\makecell{NewSQL \\ Databases}} & \cellcolor{highlighted} {TiDB v4.0~\cite{web:tidb}} & \cellcolor{highlighted}{Storage-based} & \cellcolor{highlighted}Consensus &\cellcolor{highlighted} CFT(Raft) &\cellcolor{highlighted} Concurrent &\cellcolor{highlighted} \xmark &\cellcolor{highlighted} LSM Tree(TiKV) &\cellcolor{highlighted}  \cmark(\cmark) \\

		\cdashline{2-9} \\
		 & CockroachDB~\cite{web:cockroach} & Storage-based & Consensus & CFT(Raft) & Concurrent & \xmark & LSM Tree(RocksDB) &  \cmark(\cmark) \\
		\cdashline{2-9} \\
		 & Spanner~\cite{web:spanner} & Storage-based & Consensus & CFT(Paxos) & Concurrent & \xmark & LSM Tree &  \cmark(\cmark) \\

		\cdashline{2-9} \\
		 & H-store~\cite{kallman2008h} & Storage-based & Primary-backup & CFT & Concurrent & \xmark & B Tree &  \cmark(\cmark) \\

		\midrule
		\multirow{3}{*}{\makecell{NoSQL \\ Databases}} &\cellcolor{highlighted} Etcd v3.3~\cite{web:etcd} &\cellcolor{highlighted} Storage-based &\cellcolor{highlighted} Consensus &\cellcolor{highlighted} CFT(Raft) &\cellcolor{highlighted} Serial &\cellcolor{highlighted} \xmark &\cellcolor{highlighted} B Tree(BoltDB) & \cellcolor{highlighted} \xmark(\xmark) \\

		\cdashline{2-9} \\
		& Cassandra~\cite{web:cassandra}& Storage-based & Primary-backup & CFT & Concurrent & \xmark & LSM Tree &  \cmark(\xmark) \\
		\cdashline{2-9} \\
		& DynamoDB~\cite{web:dynamodb} & Storage-based & Primary-backup & CFT & Concurrent & \xmark & B Tree &  \cmark(\xmark) \\

		\midrule
		\multirow{3}{*}{\makecell{Out-of-the \\ Blockchain \\ Databases}} & BlockchainDB~\cite{el2019blockchaindb} & \performance{Storage-based} & Consensus & \security{BFT(PoW)} & \security{Serial (in each shard)} & \security{\cmark} & \security{\makecell{LSM Tree(LevelDB)\\+MPT}} &  \cmark(\xmark) \\

		\cdashline{2-9} \\
		& Veritas~\cite{veritas} & \performance{Storage-based} & Shared log& \performance{CFT(Kafka)} & \performance{\makecell{Concurrent Execution \\ Serial Commit}} &  \security{\makecell{\cmark}} & \performance{Skip List(Redis)} &  \xmark(\xmark) \\

		\cdashline{2-9} \\
		& FalconDB~\cite{peng2020falcondb} & \performance{Storage-based} & Consensus & \security{BFT(Tendermint)} & \performance{\makecell{Concurrent Execution \\ Serial Commit}} &  \security{\makecell{\cmark }} & \security{\makecell{B Tree(MySQL)\\+Merkle Tree(IntegriDB)}} &  \xmark(\xmark) \\

		\midrule
		\multirow{3}{*}{\makecell{Out-of-the \\ Database \\ Blockchains}} & \makecell{Blockchain Relational\\
    Database (BRD)~\cite{BlockchainMeetsDatabase}} & \security{Txn-based} & \makecell{Shared log} & \makecell{\performance{CFT(Kafka)} \\ \security{BFT(BFT-SMaRt)}} & \performance{Concurrent} & \security{\cmark} & \performance{\makecell{B Tree(PostgreSQL)}} &  \xmark(\xmark) \\

		\cdashline{2-9} \\
		& ChainifyDB~\cite{schuhknecht2019chainifydb} & \security{Txn-based} & Shared log& \performance{CFT(Kafka)} & \performance{Concurrent} &  \security{\cmark} & \performance{\makecell{B Tree\\(MySQL/PostgreSQL)}} &  \xmark(\xmark) \\

		\cdashline{2-9} \\
		& BigchainDB~\cite{mcconaghy2016bigchaindb} & \security{Txn-based} & Consensus & \security{BFT(Tendermint)} & \performance{Concurrent} &  \security{\cmark} & \performance{B Tree(MongoDB)} &  \xmark(\xmark) \\

		\bottomrule
		\end{tabular}
	}
\end{table*}

\subsection{Fusion of Blockchains and Databases}
The taxonomy above provides a comprehensive description of the design space of
distributed transactional systems. This taxonomy helps in illustrating the
similarities and differences between blockchains and distributed databases. It
also serves as a principled framework for understanding the recently emerging
hybrid blockchain-database systems. In this section, we discuss how these
systems fit into the design space. We provide a deeper analysis of their
performance in Section~\ref{sec:hybrids}.




\textbf{Out-of-the-blockchain Databases.} One approach toward a hybrid design is
to start with a blockchain (or a blockchain-like system) and build database
features on top of it. Examples of this approach include
BlockchainDB~\cite{el2019blockchaindb}, Veritas~\cite{veritas}, and
FalconDB~\cite{peng2020falcondb}, which provide shared and verifiable databases
for multiple distrusting parties.
They use blockchains as an integrity-protected storage, and build other database
components on top of it.
In these systems, replication is transaction-oblivious, with duplicated states,
logs, and meta-data.
\revision{ BlockchainDB replicates storage operations and uses PoW for
consensus. It inherits the authenticated state organization from the underlying
blockchain and employs multiple blockchains for storage. Therefore, it is
amenable to sharding. However, transactions are executed sequentially within
each shard.
FalconDB and Vertias also adopt storage-based replication, but use
Tendermint~\cite{buchman2016tendermint} for consensus and Kafka~\cite{web:kafka}
as the shared log, respectively.
They use a similar optimistic concurrency control mechanism as Fabric. FalconDB
outsources the authentication task to IntegriDB~\cite{zhang2015integridb}, which
enables a light-weight client to produce a proof without holding the entire
ledger. Veritas relies on trusted verifiers for the state integrity.}

\textbf{Out-of-the-database Blockchains.} Another hybrid design approach is to
start with a database, then add blockchain features to it. Examples of this
approach include BigchainDB~\cite{mcconaghy2016bigchaindb}, Blockchain
Relational Database (BRD)~\cite{BlockchainMeetsDatabase}, and
ChainifyDB~\cite{schuhknecht2019chainifydb}. In these systems, each node has its
own database and executes transactions on its database according to a global
order achieved through consensus.
\revision{ These systems adopt the transaction-based replication model where the
ledger serves as a secure shared log that stores transactions.
The nodes execute the same sequence of transactions, but on different local
databases.
}
In particular, BRD uses PostgreSQL~\cite{postgres} and the transactions contain
invocation contexts of stored procedures. BigchainDB uses
MongoDB~\cite{web:mongodb}, thus its transactions are in JSON format.
ChainifyDB allows heterogeneous relational databases, and transactions are in
the form of standardized SQL statements.
ChainifyDB uses a Kafka broker to share logs for efficiency. In contrast,
BigchainDB uses the Tendermint consensus protocol which tolerates Byzantine
failures at the expense of performance.
BRD jointly uses Kafka~\cite{web:kafka} and BFT-SMaRt~\cite{bftsmart}, an
implementation of PBFT.
These systems inherit the concurrency support of their underlying databases,
with serializable constraints according to the ledger order. However, these
systems do not protect the local states with Merkle trees and only rely on the
integrity protection of the ledger.
Finally, these systems do not support sharding.

In summary, out-of-the-database blockchains retain many design choices of
distributed databases, as their main goal is performance. In contrast,
out-of-the-blockchain databases inherit many blockchain features, as
they are more security-driven. 
Some centralized and in-cloud databases, also learning from blockchains, 
rely on a hashed chain for verifiable transactions. 
Examples include Spitz~\cite{Spitz}, QLDB~\cite{web:qldb} and LedgerDB~\cite{yang2020ledgerdb}.
Some systems provide tailored optimizations on the ledger-like structure like LogBase~\cite{logbase}.

\subsection{\revision{Discussion}}

Table~\ref{tab:systems} summarizes some representative transactional systems and
their design choices based on our proposed taxonomy.
We only consider blockchains with generic smart contracts and NoSQL databases
with key-value data model.
We exclude permissionless blockchains from our quantitative analysis, as their
security-performance tradeoffs have been extensively
studied~\cite{gervais2016security}.
One can observe from Table~\ref{tab:systems} that the hybrid systems, just like
permissioned blockchains, share some security-oriented design choices with
blockchains and some performance-oriented design choices with databases.

\section{Experimental Setup}
\label{sec:setup}

\subsection{\revision{Systems}}

\begin{figure*}[tp] 
\centering
    \begin{subfigure}{0.8\textwidth}
		\includegraphics[width=0.99\textwidth]{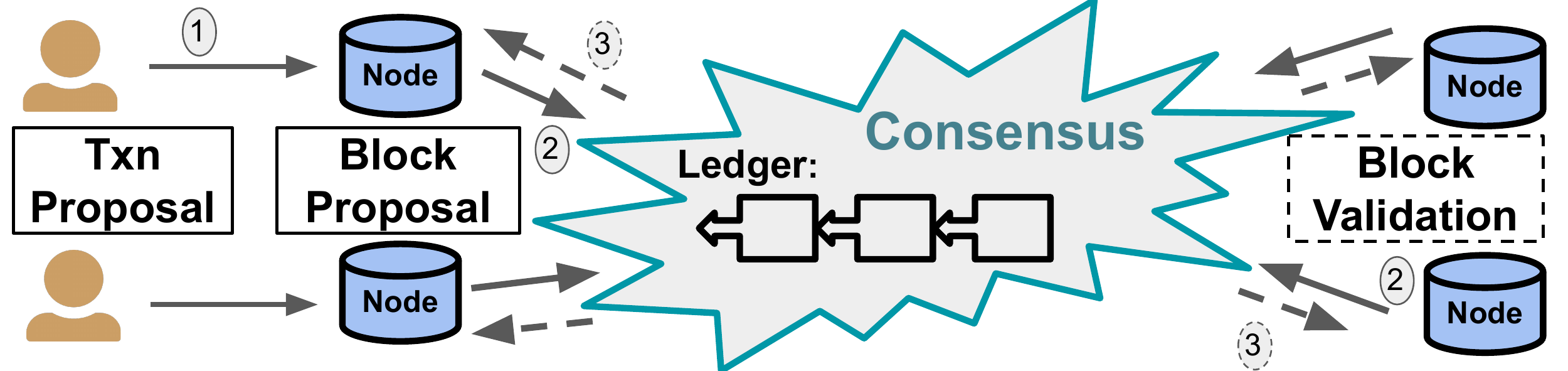}
		\caption{Quorum transaction lifecycle}
		\label{fig:quorum-arch}     
    \end{subfigure}
	\begin{subfigure}{0.8\textwidth}
		\includegraphics[width=0.99\textwidth]{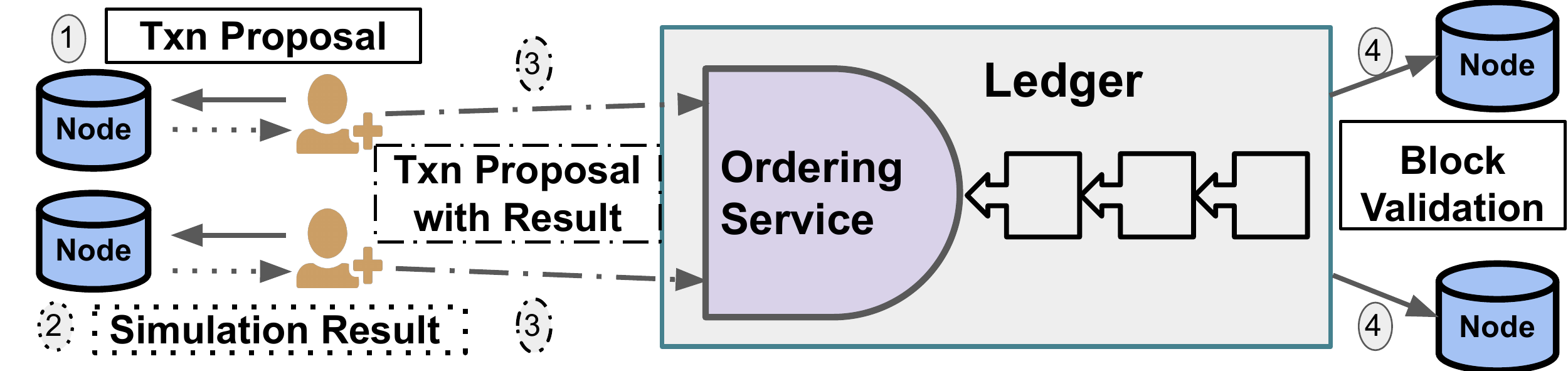}
		\caption{Fabric transaction lifecycle}
		\label{fig:fabric-arch}
	\end{subfigure}
  \caption{Transaction execution in Quorum vs. Fabric. In Quorum, a node
  assembles pre-executed transactions into blocks before running the consensus.
  In Fabric, a client collects simulation results and endorsements from peer
  nodes to form a transaction. Orderer nodes order the transactions and batch
  them into blocks, which are then pulled by the peer nodes for independent
  validation and commit.} 
 \label{fig:fabric_vs_quorum}
 \end{figure*}

We select four representative systems: two permissioned blockchains, namely Quorum~\cite{web:quorum} and Hyperledger Fabric~\cite{web:fabric}, and two distributed databases, namely TiDB~\cite{web:tidb} and etcd~\cite{web:etcd}.
Quorum represents order-execute blockchains, while Fabric represents execute-order-validate blockchains. 
They also employ different replication approaches, as shown in Figure~\ref{fig:fabric_vs_quorum}.
Fabric employs an external ordering service while Quorum relies on Raft consensus.
\revision{
Quorum is a fork of \textit{geth}, the Golang implementation of Ethereum. 
Quorum replaces the original Proof of Work (PoW) of Ethereum with a CFT protocol, namely Raft, and a BFT protocol called
Istanbul BFT (IBFT). However, it inherits Ethereum Virtual Machine (EVM) to invoke smart contracts.
}
Fabric is featured for its modularized design. 
In particular, a node role is separated into \textit{orderer} and \textit{peer}, as detailed in Figure~\ref{fig:fabric-arch}.

TiDB~\cite{web:tidb} and etcd~\cite{web:etcd} represent NewSQL and NoSQL
distributed databases, respectively.
TiDB consists of three independent modules, namely Placement Driver for
coordinating cluster management, TiKV as the replicated key-value storage, and
TiDB-server for parsing and scheduling SQL queries in a stateless manner. TiDB
only supports snapshot isolation.
Etcd provides a simple key-value data model with relaxed transactional
restrictions but focuses on the tradeoff between availability and consistency.
Similar to blockchains, etcd employs a single consensus instance to sequence all
the requests.
Without sharding, etcd fully replicates the data on each node.
We also benchmarked CockroachDB~\cite{web:cockroach}, another NewSQL database. Since it exhibits similar performance trends as TiDB, we decided to omit it in this paper.

\subsection{Setup}
For a fair comparison, we run all systems in full replication mode where each
node has a complete copy of the states. In particular, for Fabric the
endorsement policy is set such that a transaction is executed and endorsed by
all peers. For TiDB, we set the replication factor to be the same as the number
of nodes. \revision{In other words, even though TiDB partitions data to multiple
shards and manages the shards separately, each node has a copy of the entire
system state.} We configure Quorum and Fabric to use Raft, a CFT consensus.
For Fabric, we fix the number of orderers to three while scaling the peers.
For TiDB, we scale all its modules with the number of nodes.

Unless otherwise specified, we use YCSB and Smallbank workloads in our
experiments. The experiment parameters for YCSB are summarized in
Table~\ref{tab:parameter} with the default values underlined.
For the database experiments we use the open-source driver for
YCSB~\cite{web:ycsb} and the OLTPBench~\cite{difallah2013oltp} driver for
Smallbank. 
Both Fabric and Quorum are benchmarked using Caliper~\cite{web:caliper}.
We note that although there are differences in the types of drivers for
benchmarking blockchains and databases, they alone do not account for the large
performance gap reported in the following section.
 
Our experiments are conducted on an in-house cluster consisting of $96$ nodes
connected via 1Gb Ethernet.
Each node is equipped with Intel Xeon E5-1650 CPU, 32GB RAM, and 2TB hard disk.
All the experiments are repeated three times and we report the average.

\begin{table}
	\centering
	\caption{Experiments parameters}
	\label{tab:parameter}
	\begin{tabular}{@{}ll@{}}
	\toprule
	\textbf{Variable}               & \textbf{Values}               \\
	\midrule
	Record size (Byte)               & 10, 100, \underline{1000}, 5000          \\
	Zipfian coefficient $\theta$       & \underline{0.0}, 0.2, 0.4, 0.6, 0.8, 1.0 \\
	\# of transaction operations & \underline{1}, 2, 4, 6, 8, 10            \\
	\# of nodes & 3, \underline{5}, 7, 11, 15, 19            \\
	\bottomrule
	\end{tabular}
\end{table}

\section{Result and Analysis}
\label{sec:result}
We first summarize the main findings, then provide the detailed experimental analysis. Based on these findings,
we propose an empirical framework that compares the performance of recent hybrid blockchain-database systems.
The framework not only explains the performance differences in existing systems, but it is also useful for
understanding future hybrid systems.

\begin{itemize}
  \item \textbf{Peak performance.} The performance gap between blockchains and distributed databases is large. However, the gap is not as significant as previously reported.
  \item \textbf{Replication. }
  \revision{
    The transaction-based replication model restricts concurrency, which limits the impact of different
	replication approaches and failure models on the system's peak performance. 
  }
  \item \textbf{Concurrency.} Execute-order-validate blockchains have low performance under workloads with high contention and constraints.
  \revision{
  The impact of workloads on the performance is prominent in NewSQL databases, where concurrency is
  on top of replication.}
  \item \textbf{Storage. } The ledger abstraction in blockchains incurs significant storage overhead. On the
  other hand, the overhead needed to guarantee state tamper evidence is small.
  \item \textbf{Sharding. } The performance of sharded blockchains is far behind that of distributed
  databases, due to the security requirements on shard formation and periodic reconfiguration.
\end{itemize}

\subsection {Peak Performance}
\subsubsection{YCSB}

\begin{figure*}[tp]
	\centering
    \begin{subfigure}{0.4\textwidth}
        \includegraphics[width=0.99\textwidth]{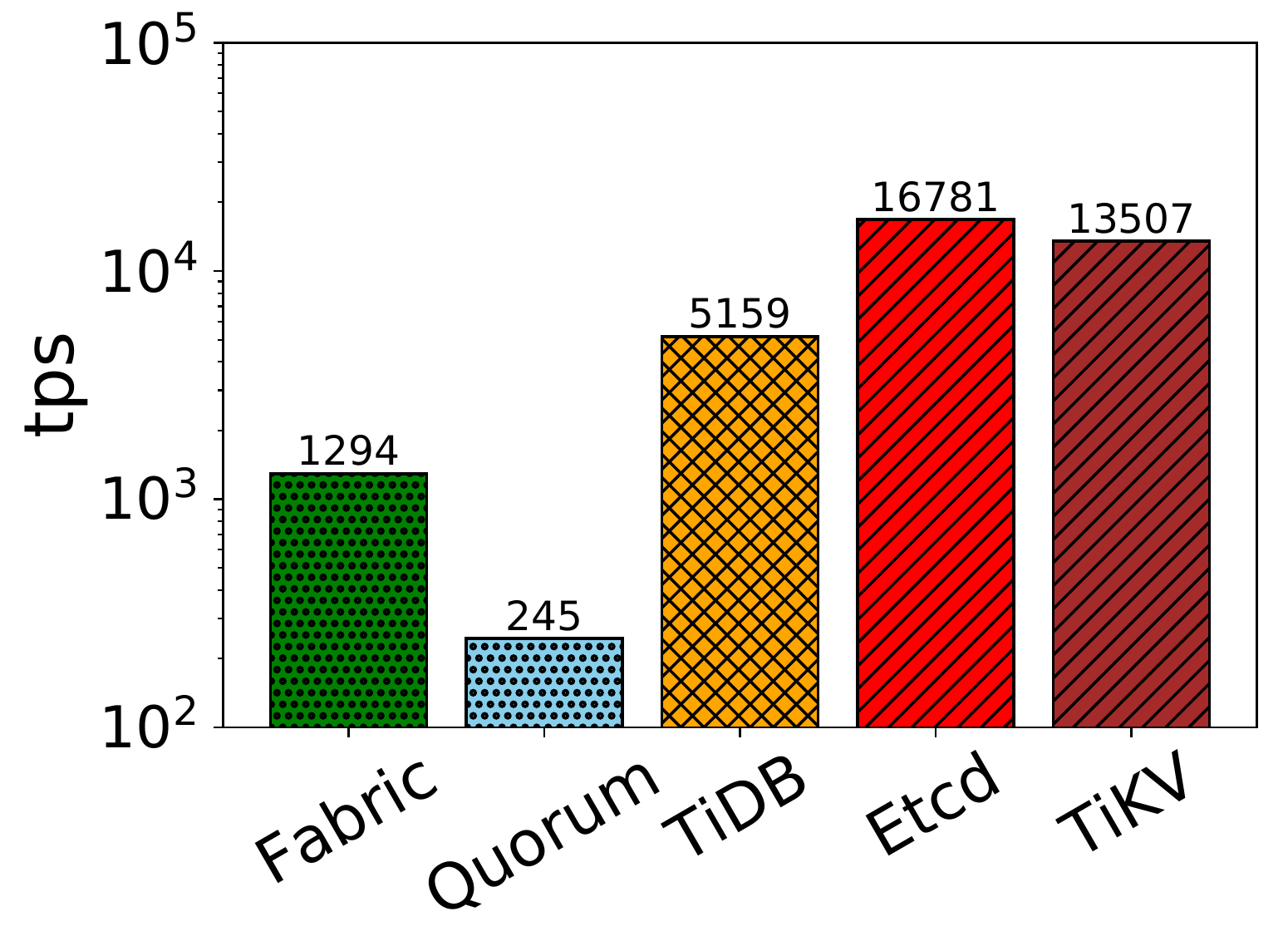}
        \caption{Update}
        \label{fig:ycsb-update-thruput}
    \end{subfigure}
    \begin{subfigure}{0.4\textwidth}
        \includegraphics[width=0.99\textwidth]{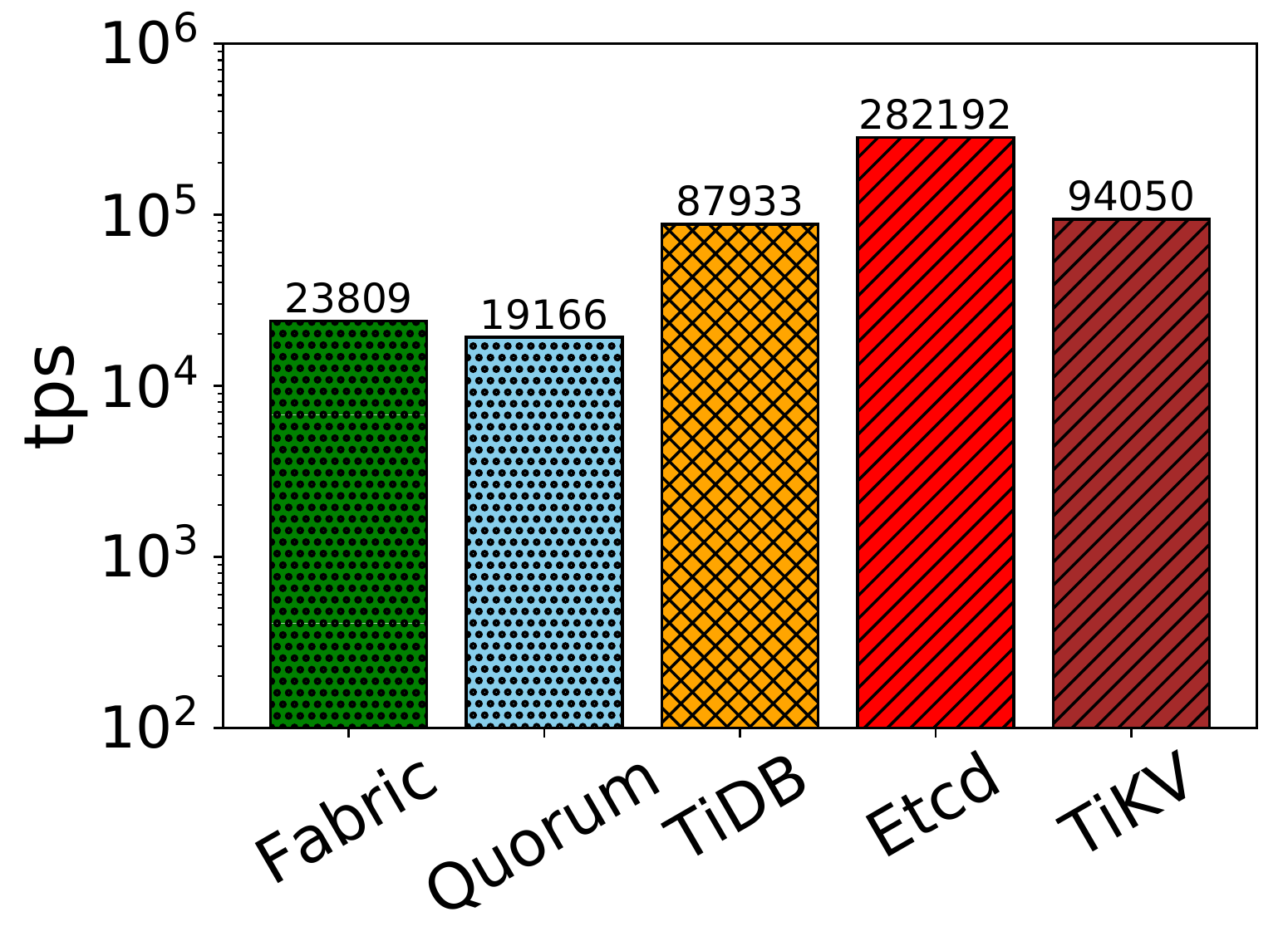}        
        \caption{Query}
        \label{fig:ycsb-query}
    \end{subfigure}
    \caption{Throughput of YCSB workload (log scale).}
    \label{fig:ycsb-thruput}
\end{figure*}

We first analyze the peak performance of the four systems under the default
configurations shown in Table~\ref{tab:parameter}.
Specifically, we populate each system with 100K records, each of size $1$ KB.
We then measure the throughput and latency against two YCSB workloads: uniform
update-only (100\% writes) and uniform query-only (100\% reads). We also measure
independently the performance of TiKV, the replicated storage of TiDB, and
include it in this comparison.

Figure~\ref{fig:ycsb-thruput} shows the peak throughput of the five systems.
The relational NewSQL database (TiDB) outperforms the blockchains, while the
replicated storages (etcd and TiKV) outperform the relational database.
Specifically, the two blockchains achieve update throughputs of below $1500$
transactions per second (tps), whereas TiDB achieves $5159$ tps.
The two key-value storages, etcd and TiKV, achieve around $15,000$ tps.
Both outperform the NewSQL database because they do not incur the overhead of
supporting ACID transactions.
This is evidenced by the gap between TiDB and TiKV, caused by the overhead of
the TiDB-server that wraps around the key-value storage.
But this gap is less evident under the query workload, as ACID semantics impose
less constraints on read-only transactions.

\begin{figure*}[tp]
	\centering
	\begin{subfigure}{0.4\textwidth}
		\includegraphics[width=0.99\textwidth]{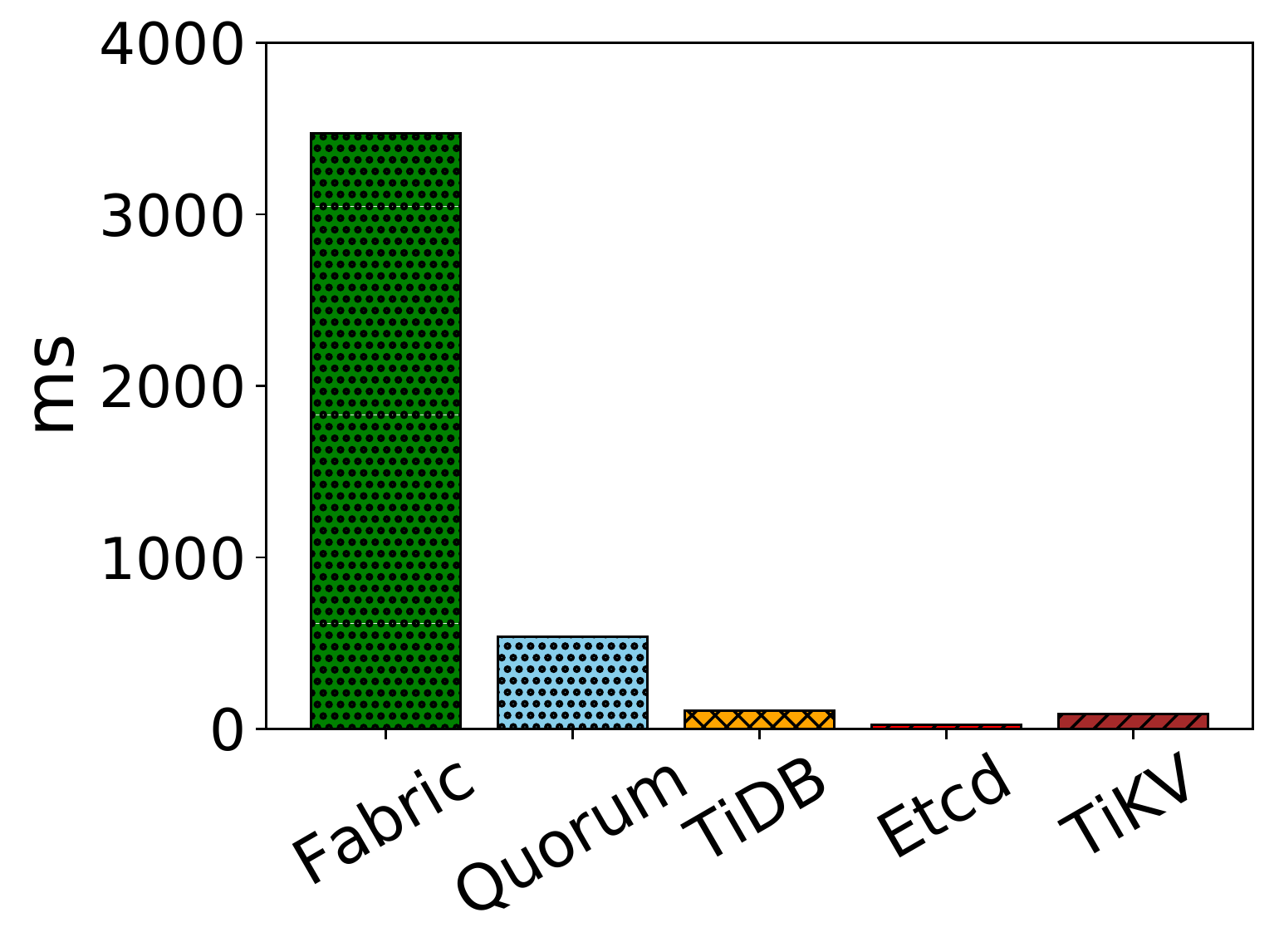}
        \caption{Update}
		\label{fig:ycsb-update-delay}
	\end{subfigure}
	\begin{subfigure}{0.4\textwidth}
		\includegraphics[width=0.99\textwidth]{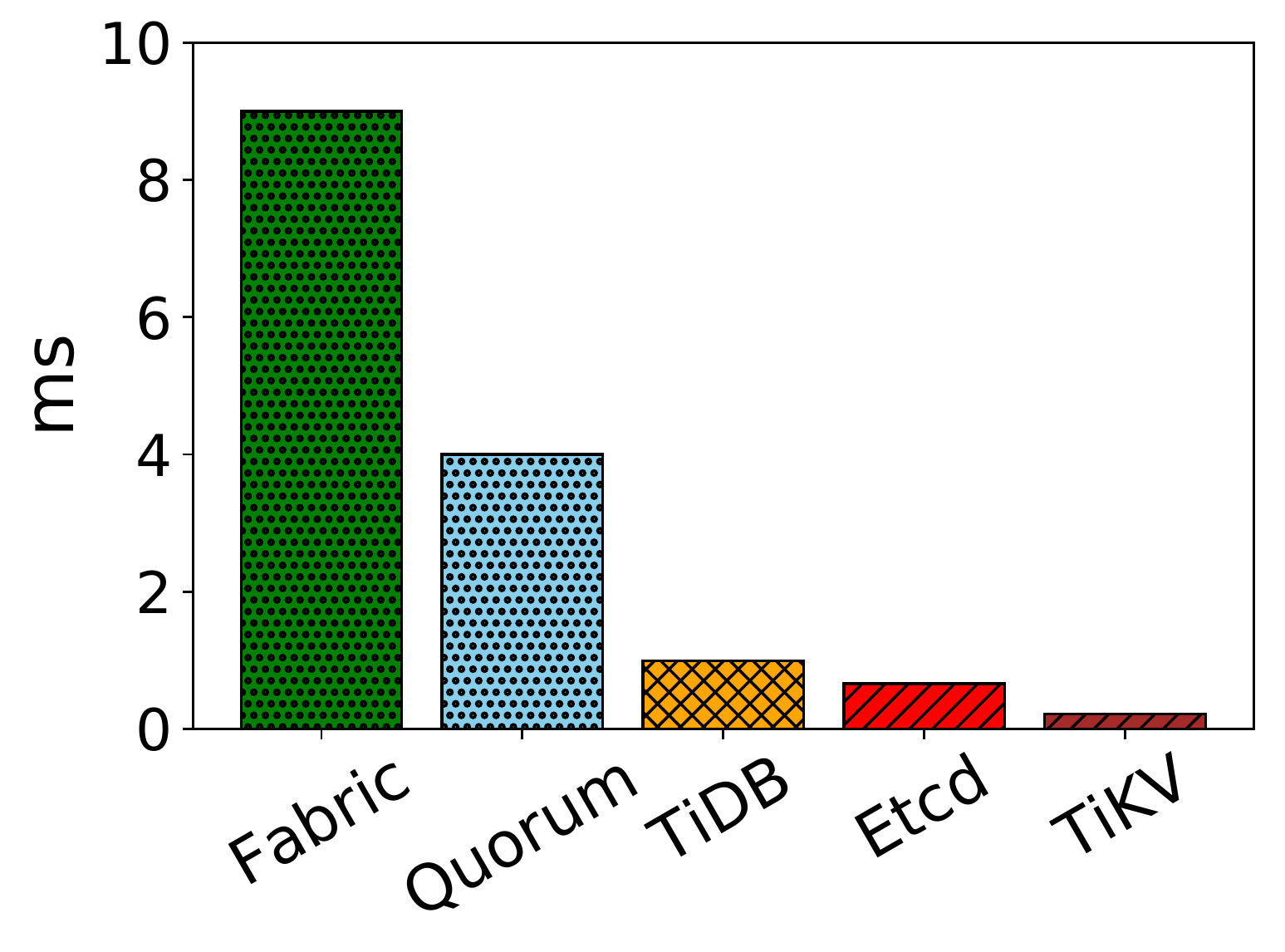}		
        \caption{Query}
		\label{fig:ycsb-query-delay}		
	\end{subfigure}
	\caption{Latency of YCSB workload.}
	\label{fig:ycsb-delay}
\end{figure*}

Figure~\ref{fig:ycsb-delay} shows the latency when the systems are unsaturated.
Similar to throughput, we observe a clear separation between the blockchains and
the databases.
We note that the blockchains have weaker guarantees for read-only transactions
compared to those offered by the databases (linearizability).
Responses to read requests in the former still take longer (up to $6\times$ in
Fabric) than the linearizable reads in the latter.
The update (query) latency in Fabric and Quorum is around $3500$ms ($9$ms) and
$500$ms ($4$ms) respectively, while in databases it is below $100$ms ($1$ms).

Our results confirm the conclusion drawn in~\cite{dinh2017blockbench} that the
performance of blockchains lags far behind state-of-the-art databases.
However, we observe a smaller gap than that reported
in~\cite{dinh2017blockbench}.
In particular, the relational database, TiDB, achieves $4\times$ greater
throughput than the fastest blockchain, Fabric, under the uniform update
workload (5159 vs.
1294 tps).
This is in contrast with~\cite{dinh2017blockbench}, where H-Store exhibits more
than $120\times$ speedup over blockchains.
The key reason is that H-Store is an in-memory, distributed database with
primary-backup replication.
H-Store represents an extreme point of the design space that makes it rather
dissimilar to blockchains.
In contrast, all systems considered in our work incur some overheads from the
consensus protocols.

\begin{figure*}[tp]
	\begin{minipage}{0.4\textwidth}
		\centering
        \includegraphics[width=0.99\textwidth]{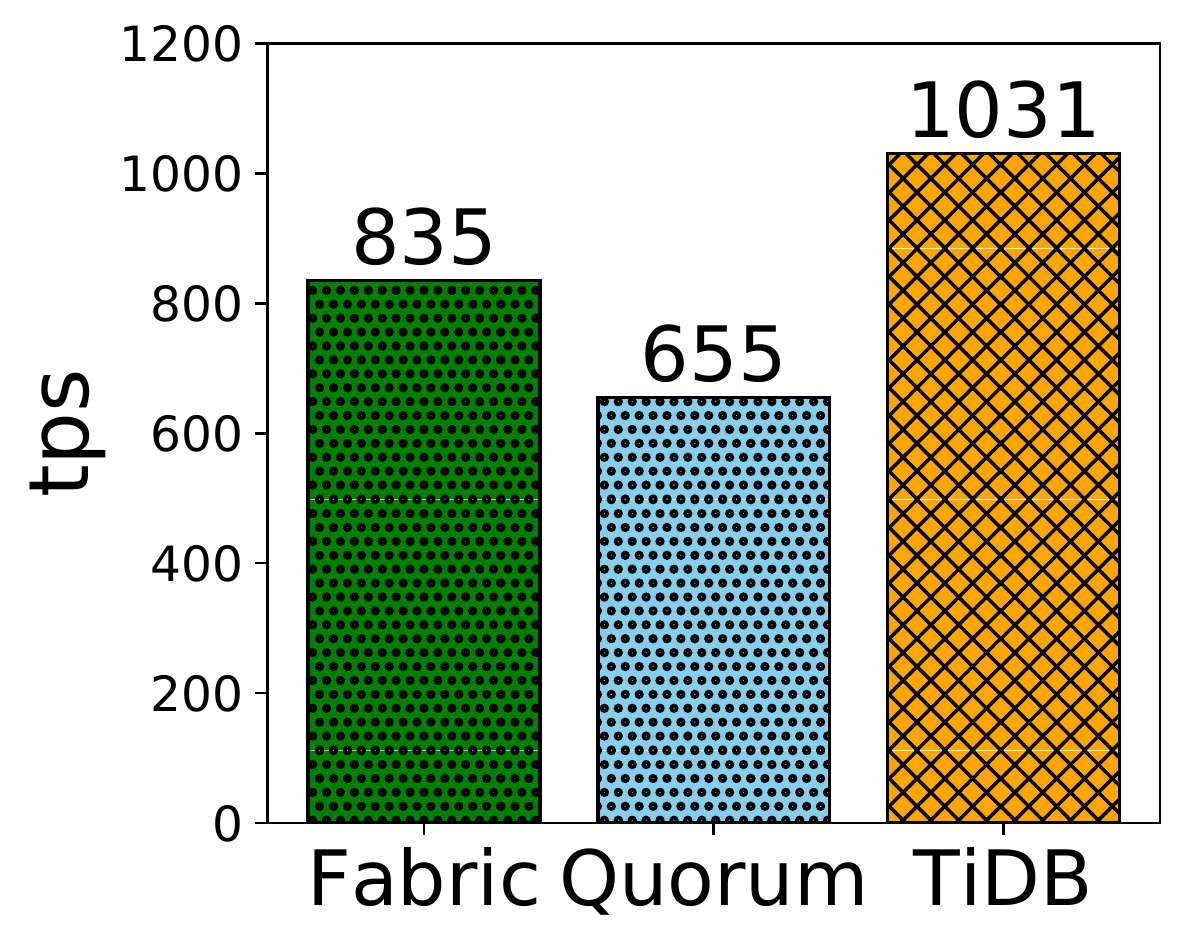}
		\caption{Throughput of the skewed Smallbank workload (1M records).}
		\label{fig:sb} 
	\end{minipage} \hfill
	\begin{minipage}{0.4\textwidth}
		\centering
		\includegraphics[width=0.99\textwidth]{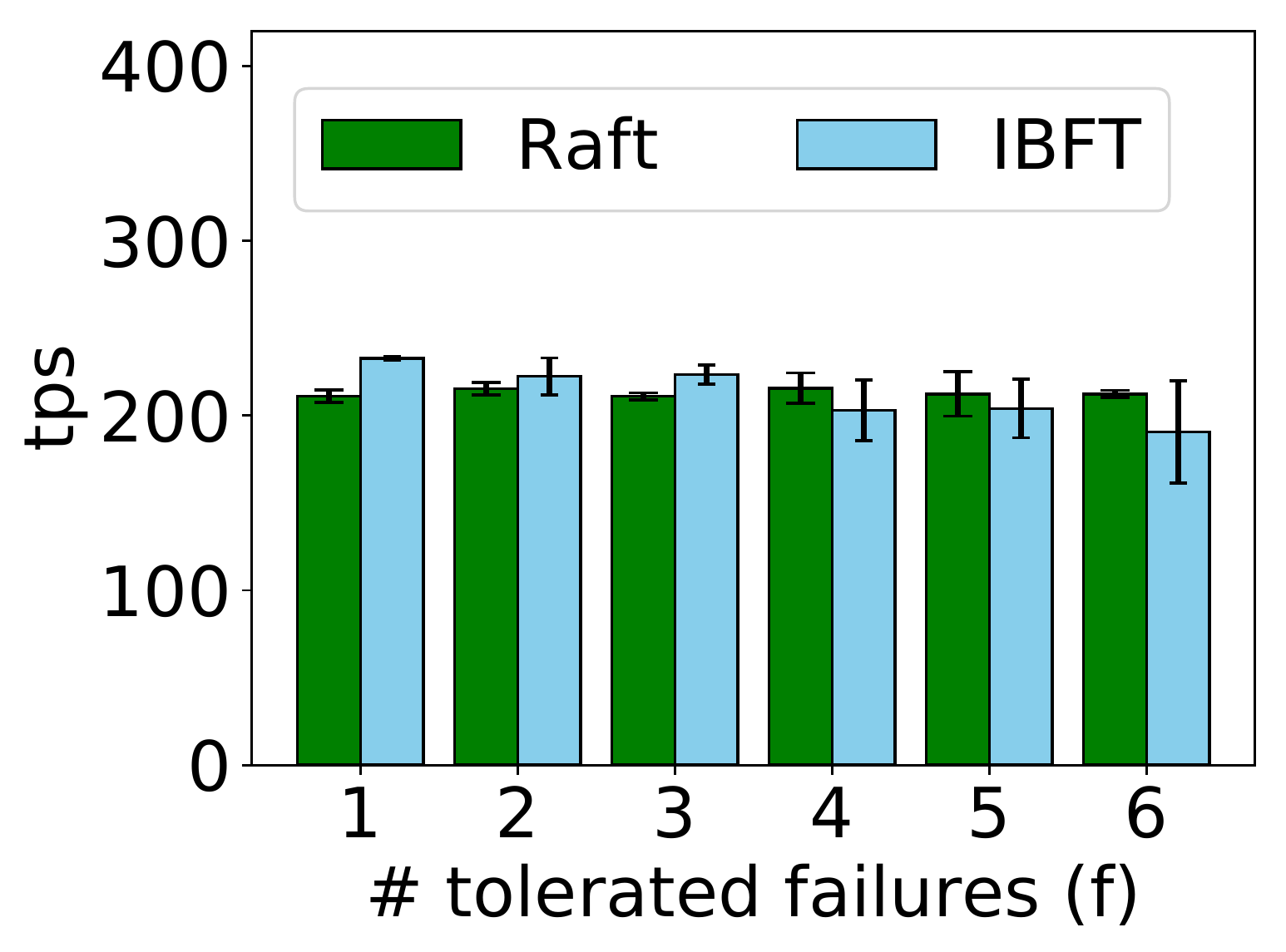}
		\caption{Quorum throughput with CFT(Raft) and BFT(IBFT).}
		\label{fig:quorum-consensus}
	\end{minipage} \hfill
	\begin{minipage}{0.93\textwidth}
			\centering
			\begin{subfigure}{0.49\textwidth}
				\includegraphics[width=0.99\textwidth]{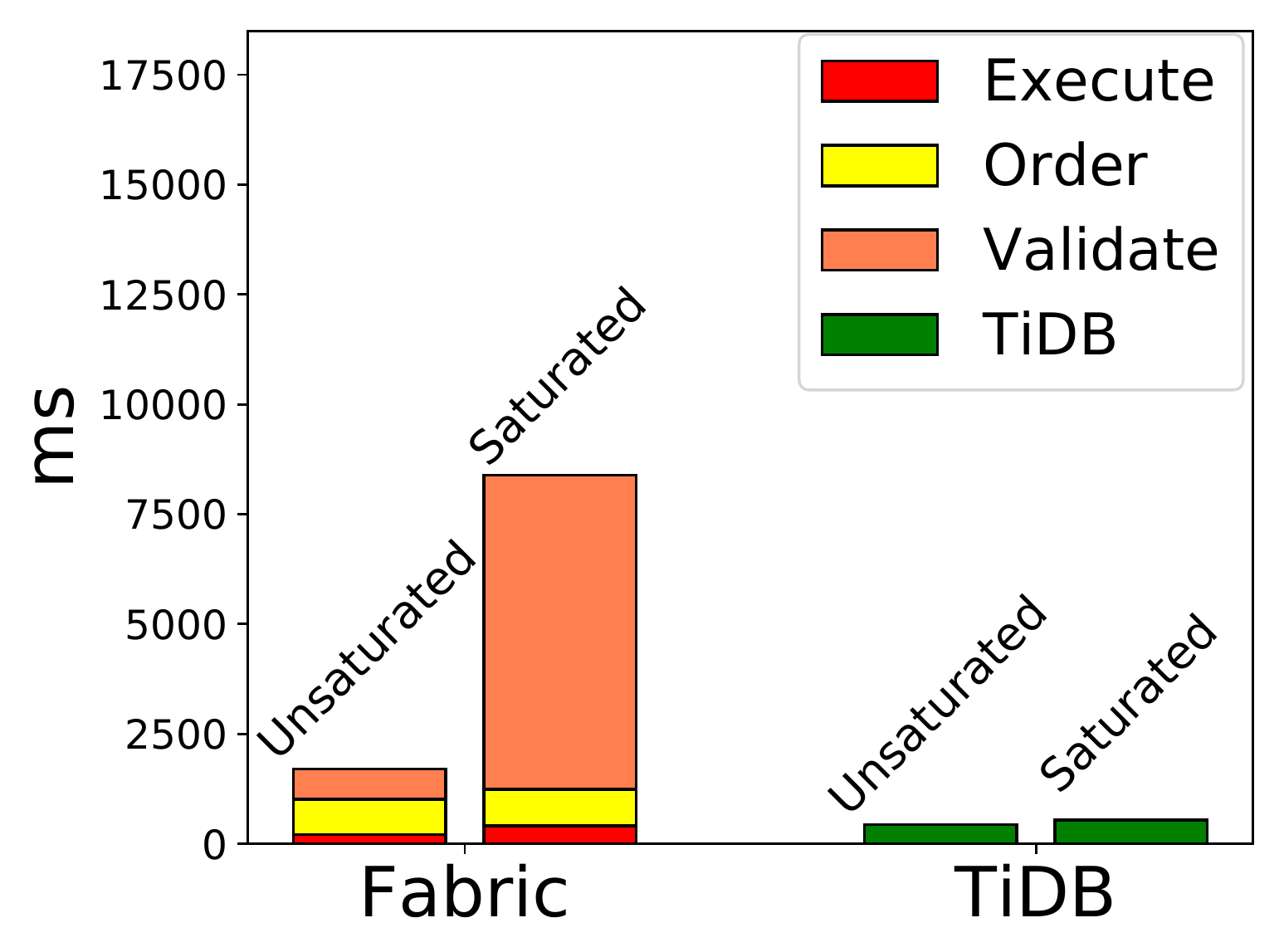}
				\caption{Update}
				\label{fig:update-breakdown}
			\end{subfigure}
			\begin{subfigure}{0.49\textwidth}
				\includegraphics[width=0.99\textwidth]{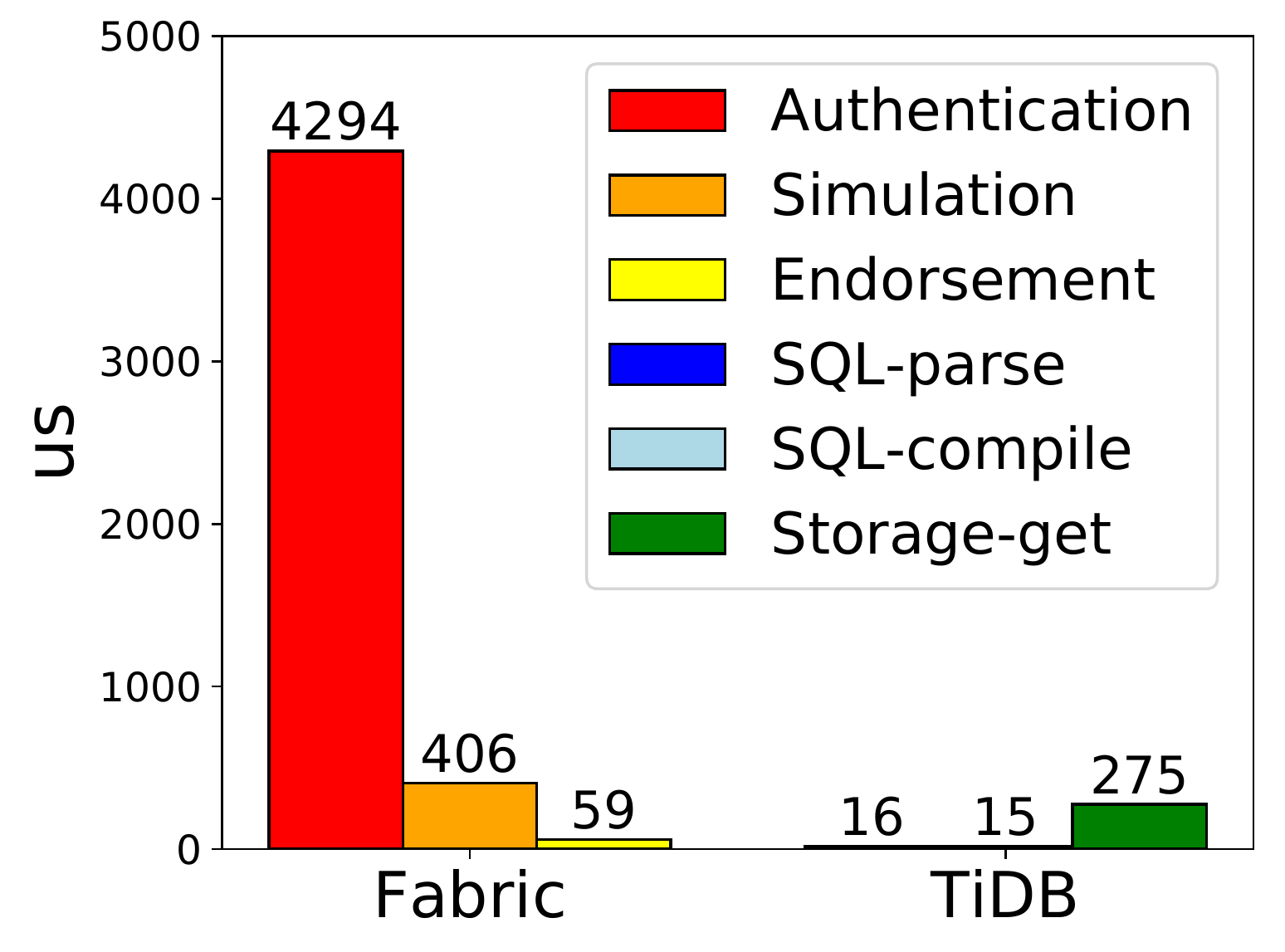}
				\caption{Query}							
				\label{fig:query-breakdown}		
			\end{subfigure}
			\caption{Latency breakdown.}
	\end{minipage} \hfill
\end{figure*}

\subsubsection{\revision{Smallbank}}

Figure~\ref{fig:sb} compares the OLTP performance under the Smallbank workload.
The request key follows a Zipfian distribution with coefficient $\theta=1$ on 1M
records.
We do not include etcd because it does not support general transactional
workloads.
Compared to YCSB, besides skewness, a Smallbank transaction imposes more
constraints and may touch up to two records, but the record size is smaller.
To our astonishment, the experiments show that the performance difference
between blockchains and distributed databases is small. For example, Fabric and
Quorum exhibit throughputs of $835$ and $655$ tps, respectively, while TiDB
exhibits only $1031$ tps.
The performance of Fabric and TiDB drops when switching from YCSB to Smallbank,
while the performance of Quorum improves with a peak throughput under Smallbank
that is $2.5\times$ greater compared to YCSB.
We attribute this improvement to the smaller record size of Smallbank.
As we shall see in Section~\ref{sec:exp:txn:record_size}, Quorum's performance
is vulnerable to transactions that access large records.
Likewise, the request skewness accounts for the throughput drop reported by Fabric and TiDB. 

\subsection{Replication}


\subsubsection{Effect of replication model}
\label{sec:exp:replication:model}
To understand the impact of the replication model, we focus on Fabric and TiDB
because they support different transaction lifecycles.
Figure~\ref{fig:update-breakdown} compares the latency of a transaction when the
systems are both unsaturated and saturated. Besides its higher latency compared
to TiDB, Fabric exhibits a significant increase in latency when the system is
saturated. To investigate this issue, we instrument Fabric codebase to record
detailed latency breakdown at each phase of a transaction. In particular, we
measure the latency of the execute, order, and validate phases.
When Fabric is unsaturated, the order and validate phases take roughly $700$ms
each, while the execute phase takes below $500$ms.
But when the request rate exceeds the system capacity, validation phase
becomes the bottleneck, as shown in Figure~\ref{fig:update-breakdown}.


We attribute this increase in latency to the serial validation of
blocks in Fabric, where blocks pile up before committing their transactions.
Even inside a block, transactions persist their effects sequentially
based on their internal order.
Worst still, substantial overhead in transaction processing is attributed to
factors other than data processing.
For example, we observe that Fabric, under the saturated scenario, spends $42\%$
of the block validation time to verify the transaction signature.
We note that serial validation is Fabric's implementation choice, i.e., it could
commit transactions concurrently.
However, most of the blockchains impose a strict transaction order to achieve
deterministic execution for security reasons.
In contrast, database transactions do not suffer from such strict sequentiality
under their storage-based replication, nor do they incur security overhead.

The security overhead is the most prominent in query transactions,
which involve no consensus in both systems.
We show in Figure~\ref{fig:query-breakdown} that Fabric spends most of the query
time to authenticate the clients.
In contrast, TiDB incurs no cryptographic overhead and most of its query time is
spent on getting the data.


\subsubsection{\revision{Effect of replication approach}}
\label{sec:exp:replication:approach}

\begin{table}[tp]
    \centering
    \caption{Throughput (in tps) with varying number of nodes under full replication mode.}
    \begin{tabular}{@{}lrrrrr@{}}
    \toprule
    \textbf{} & \textbf{3} & \textbf{7} & \textbf{11} & \textbf{15} & \textbf{19} \\ \midrule
    Fabric             & 1560        & 1288        & 1031         & 749         & 528         \\
	Quorum             & 237        & 236        & 229         & 217         & 219         \\
    TiDB               & 5697       & 7884       & 7544        & 6239        & 5526        \\
    Etcd               & 19282      & 16453      & 11243        & 7801        & 6076        \\ 
	\bottomrule
    \end{tabular}
    \label{tab:scale}
\end{table}

\begin{table}[tp]
    \centering
    \caption{Throughput (in tps) when independently varying the number of TiDB servers and TiKV nodes in a cluster under full replication mode.}
    \begin{tabular}{@{}c|rrrrr@{}}
    \toprule
    \backslashbox{TiKV}{TiDB} & \textbf{3} & \textbf{7} & \textbf{11} & \textbf{15} & \textbf{19} \\ \midrule
	\textbf{3}	& 5697	& 8517	& 9116	& 8838	& 8690	\\
	\textbf{7}	& 5951	& 7884	& 8539	& 8162	& 8246	\\
	\textbf{11}	& 5847	& 6871	& 7544	& 6941	& 7429	\\
	\textbf{15}	& 5121	& 5703	& 6306	& 6239	& 5618	\\
	\textbf{19}	& 4198	& 5238	& 5477	& 5563	& 5526	\\
	\bottomrule
    \end{tabular}
    \label{tab:tidb}
\end{table}

We increase the number of nodes to compare the scalability of shared log and
consensus-based systems under the full replication mode, and summarize the
results in Table~\ref{tab:scale}.
Here, Fabric is the only shared log system.
Even though Fabric employs the Raft consensus to obtain the transaction order,
this is an external service with 3 fixed orderers. 
The increasing number of Fabric peers consume the same shared ordered log while for the other systems, all their nodes participate in the consensus.


Contrary to our expectations on the two blockchains, we observe neither a constant performance of the
shared log system nor performance degradation in the consensus-based system.
In particular, Fabric's throughput drops $3\times$ from 3 to 19 nodes, while
Quorum's throughput is roughly unchanged.
In Fabric, we find a $38\%$ increase in the block validation latency.
This is because the endorsement policy requires a transaction to be endorsed by
all the nodes.
Hence, more nodes lead to transactions with more signatures and, therefore,
longer validation.
Due to the sequentiality in transaction-based replication, this increase in
validation time translates to the decrease in throughput, as we explained in
Section~\ref{sec:exp:replication:model}.
On the other hand, Quorum underutilizes Raft, making its performance insensitive
to the consensus group size.
Specifically, Quorum first pre-executes transactions at the tip of the ledger,
before batching these transactions into a block for the consensus.
Thus, the block proposal rate is affected by the ledger's sequentiality.

\revision{Under the same Raft protocol, the NoSQL database, etcd, achieves
higher peak performance compared to the blockchains, but the performance
degrades with the number of nodes.
We attribute this to the consensus protocol.}
The NewSQL database does not exhibit either a constant or decreasing performance
trend.
Instead, TiDB reaches its peak performance on 7 nodes.
This is because of the interplay between the TiKV storage and the
transaction processing on TiDB servers.
To study this interplay, we vary the number of TiDB servers and TiKV nodes independently and report the throughput in Table~\ref{tab:tidb}.
The TiDB servers, due to their stateless nature, should exhibit a linear speedup with more number of nodes~\cite{pingcap}. 
However, we observe that when the number of TiDB servers is low, they become the bottleneck since there in not enough processing capacity for the SQL statements. 
But when the number of TiDB servers increases, TiKV becomes the bottleneck.
We identify two distinct factors that affect the performance of TiKV.
On the one hand, a larger consensus group in TiKV leads to a greater overhead under the full replication mode. 
On the other hand, under full replication, there are more TiKV nodes that can serve the requests and thus alleviate hot spots~\cite{pingcap}.
When fixing the number of TiDB servers, we notice a degrading performance along with more TiKV nodes.
This implies that the consensus overhead outweighs the gain of hot spot alleviation in TiKV.

%

Finally, we conclude that the transaction-based replication model has an obvious
impact on the performance of blockchains, while replication approaches have plain
effects on the performance of distributed databases.


\subsubsection{Effect of failure model}
\label{sec:exp:replication:consensus}
We compare the performance of Raft and Istanbul Byzantine Fault Tolerant (IBFT)
consensus in Quorum to illustrate the impact of different failure models.
Recall that Raft tolerates only crash failures, whereas IBFT can tolerate
Byzantine failures.
\revision{IBFT shares the crux of PBFT, which consists of a three-phase commit.
But IBFT is heavily optimized for blockchains.
For example, by embedding the consensus meta-data in the ledger, IBFT saves PBFT
checkpointing efforts.
IBFT additionally accommodates dynamic validators, while PBFT assumes fixed
membership.
}


Figure~\ref{fig:quorum-consensus} shows similar peak throughputs that remain
relatively constant when increasing the number of tolerated failures.
However, we observe that IBFT's throughput exhibits higher variance in larger
networks, as evidenced by the greater error bar. This is due to the larger
quorums needed in IBFT, which are $2f+1$ out of $3f+1$, compared to $f+1$ out of
$2f+1$ replicas needed in Raft.
IBFT needs to contact more replicas in a time window compared to Raft to avoid
the view change, during which the corresponding transaction processing is
interrupted.
When $f$ increases, the probability of such interruption increases accordingly,
hence, this leads to larger variances in performance.

\subsection{Concurrency} 
\label{sec:exp:concurrency}

\subsubsection{Effect of skewness}

\begin{figure*}[tp]
	\centering
	\begin{subfigure}{0.4\textwidth}
		\includegraphics[width=0.99\textwidth]{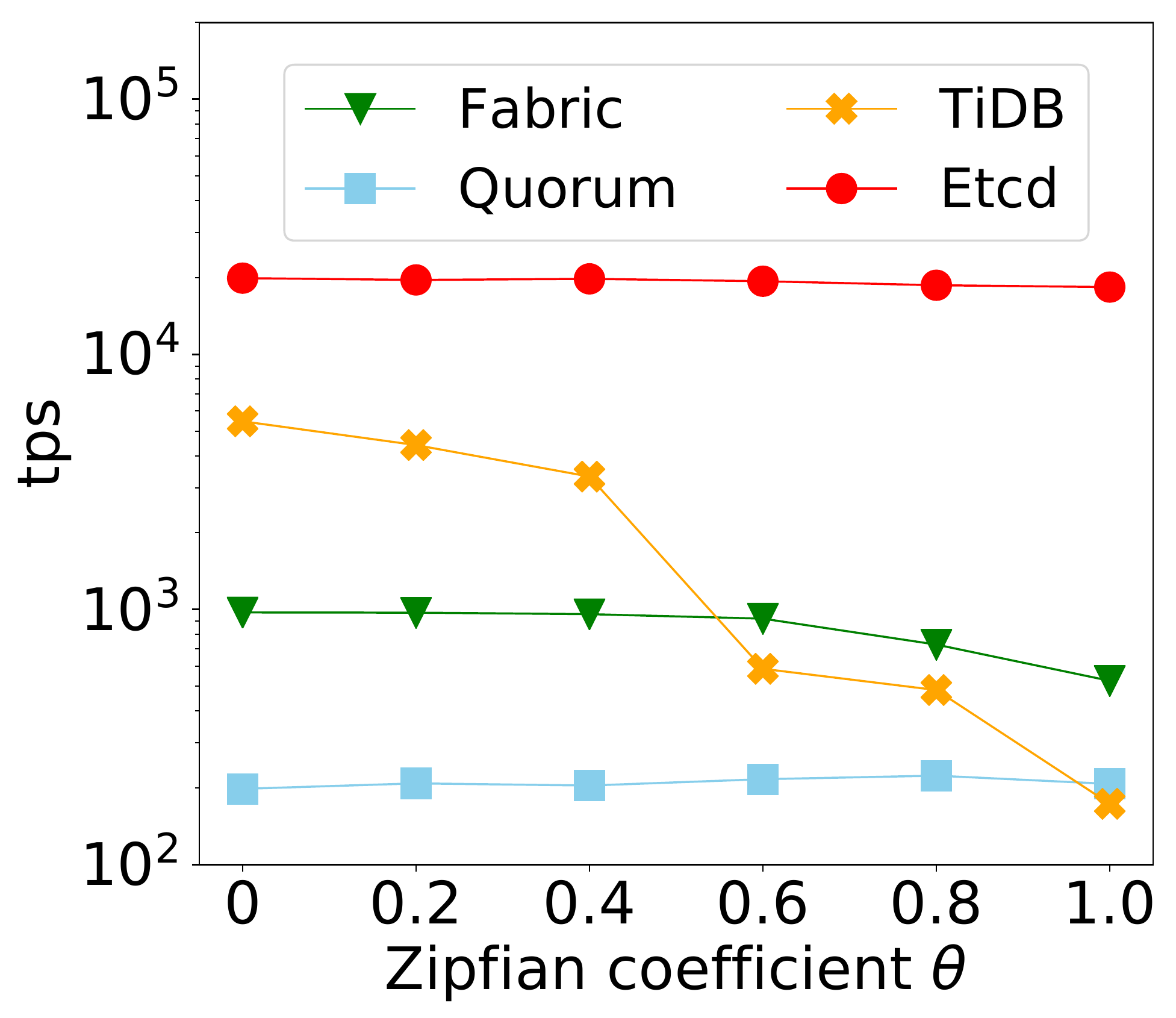}
		\caption{Throughput (log scale)}        
		\label{fig:skew-thruput}
	\end{subfigure}
	\begin{subfigure}{0.4\textwidth}
		\includegraphics[width=0.99\textwidth]{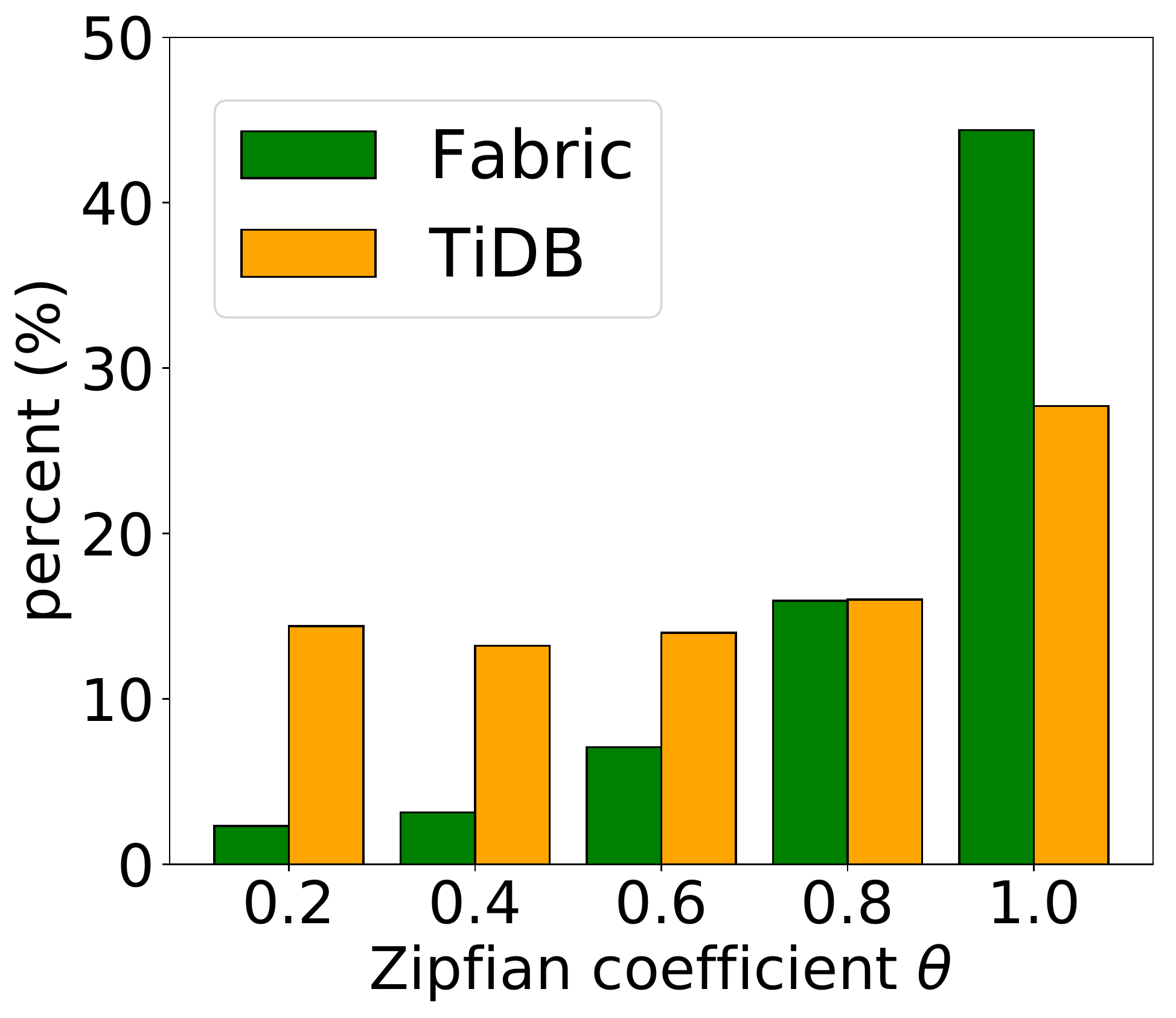}
		\caption{Abort rate}        
		\label{fig:skew-abort}
	\end{subfigure}
	\caption{Throughput and abort rate with skewed workloads. Each transaction modifies a single record.}
	\label{fig:skew}
\end{figure*}

To analyze the effect of concurrency control mechanisms, we use skewed workloads
in which each transaction modifies (first read, then update and write back) a
single record.
The records' keys follow a Zipfian distribution that varies based on the
skewness coefficient $\theta$.
Figure~\ref{fig:skew} shows the throughputs and the corresponding abort rates
under different skewness.
Our key observation here is that blockchains and databases are comparable under
a high contention workload, given the fact that TiDB drastically drops from
$5461$ to $173$ tps when $\theta$ increases from $0$ to $1$.
Etcd and Quorum do not have concurrency control because they execute
transactions serially. Thus, their performance is not affected by skewness.


Although Fabric commits transactions sequentially, we observe a $31\%$ drop in
throughput from a uniform to a skewed workload with $\theta=1$.
This is due to Fabric's optimistic concurrency control on read-write conflicts.
That is, a transaction contains the versions of the records read during the
proposal phase, which are then checked in the validation phase.
If the versions are not the latest, the transaction aborts.
A skewed workload means that many transactions are accessing the same records,
leading to a higher probability of transaction abort.
For example, Figure~\ref{fig:skew-abort} shows that $44\%$ of the transactions
in Fabric abort when $\theta=1$.

Another interesting observation is that TiDB's throughput drop is
disproportional to its increase in abort rate.
Specifically, when $\theta=1$, only $30\%$ of TiDB's transactions fail but the
throughput decreases by $90\%$.
This is because each transaction coordinator must obtain a latch on a primary
record, whose write outcome determines the overall transaction status.
But write must undergo the consensus for replication.
Under a highly skewed workload, such a latching mechanism makes the transaction
coordinator spend more time on contention resolution than the actual execution
of the transaction payload, resulting in a remarkable decrement of the overall
throughput~\cite{pingcap}.
Hence we conclude that the workload skewness exerts a tremendous impact on
storage-based replicated, concurrency-over-replication architectures.

\subsubsection{Effect of operation count}

\begin{figure*}[tp]
	\centering
	\begin{subfigure}{0.4\textwidth}
		\includegraphics[width=0.99\textwidth]{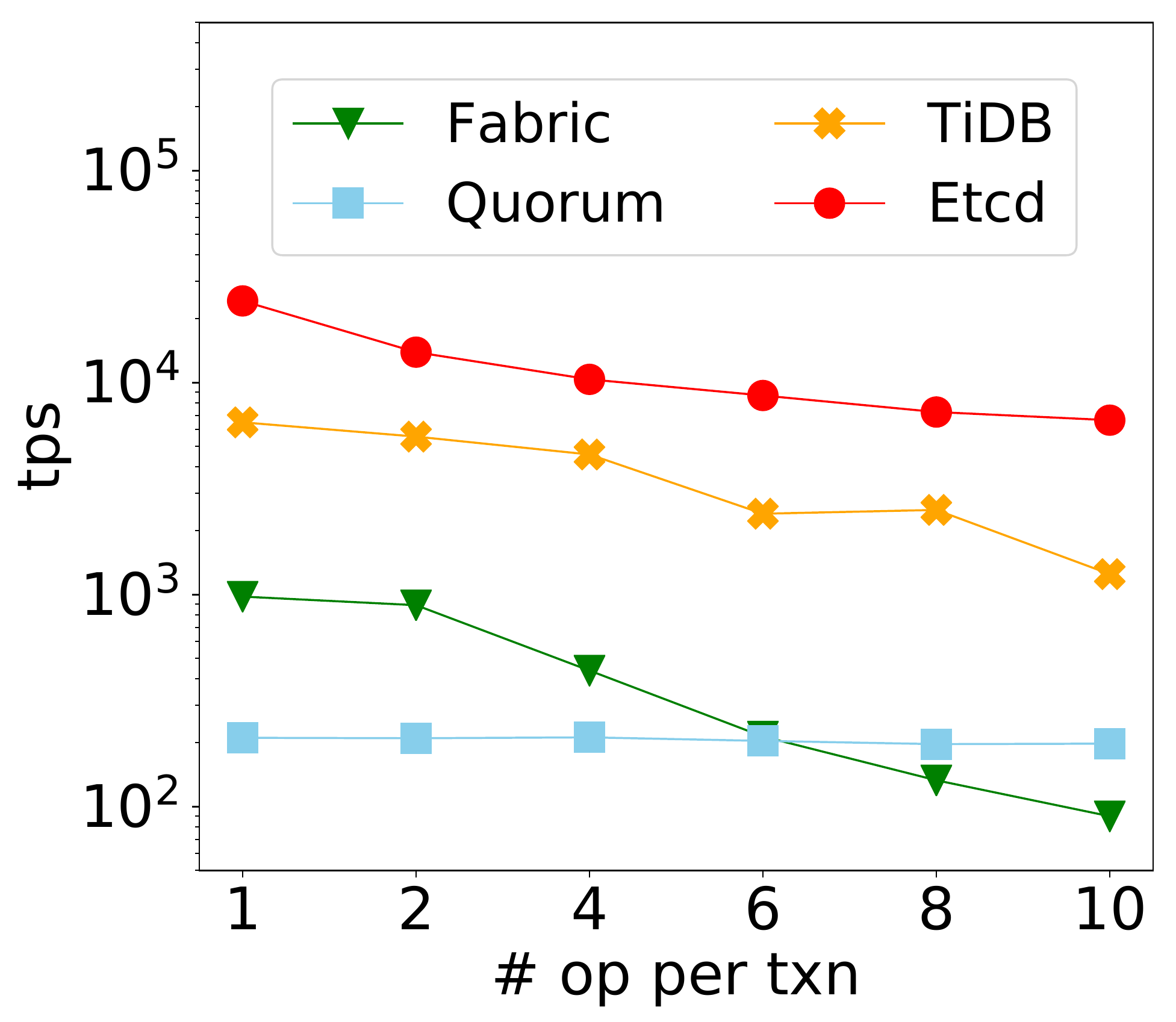}
		\caption{Throughput (log scale)}        
		\label{fig:txn-size}
	\end{subfigure}
	\begin{subfigure}{0.4\textwidth}
		\includegraphics[width=0.99\textwidth]{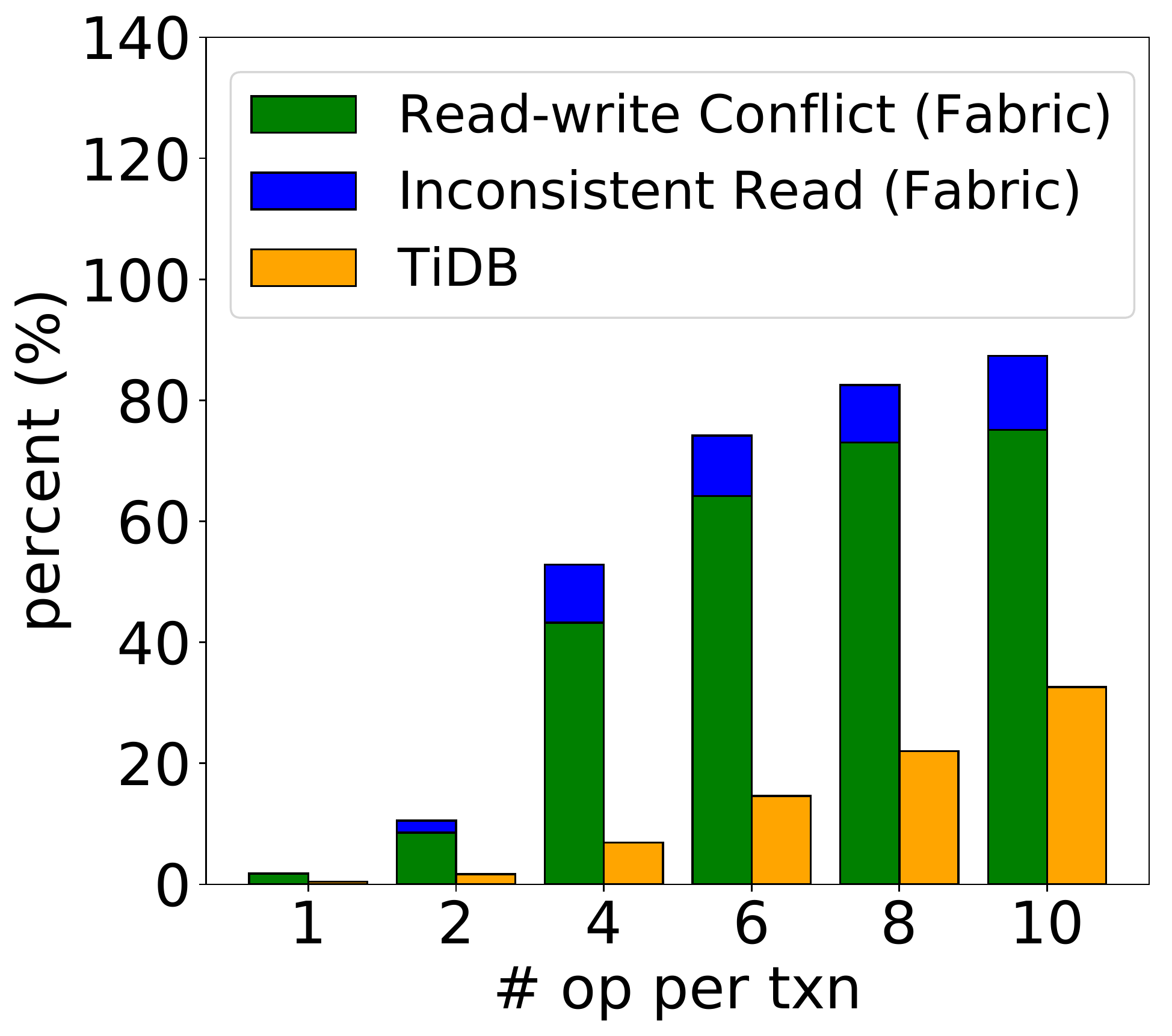}
        \caption{Abort rate}
		\label{fig:txn-size-abort}
	\end{subfigure}
	\caption{Throughput and abort rate with uniformly modified records in a single
	transaction.}
	\label{fig:txn-size-fig}
\end{figure*}

We gradually include more update operations per transaction to analyze the
impact of transaction atomicity on performance. 
To remove the effect of transaction size, for a given number of operations we
vary the record size such that the total transaction size is $1000$ bytes. 
For example, if a transaction writes 10 records, then each record contains
100 bytes.

As shown in Figure~\ref{fig:txn-size}, the performance of Fabric, TiDB, and etcd
drops when the number of operations per transaction increases.
In particular, with 10 operations per transaction, TiDB achieves only $32\%$ of
the throughput of single operation transactions.
Two sources of overheads contribute to this drop in performance.
First, there are more conflicts when a transaction writes to more records, which
leads to a higher abort rate.
Second, TiDB uses sharding, which means that a 10-operation transaction may span
multiple shards.
As there are more shards, the overhead of the 2PC coordination in TiDB
increases.
Etcd and Quorum are unaffected because they do not entail cross-shard
transactions.

Figure~\ref{fig:txn-size-abort} shows the abort rate of TiDB and Fabric as the
number of operations per transaction increases.
Both systems experience high abort rates: $26.9\%$ for TiDB and $87\%$ for
Fabric.
Interestingly, while TiDB aborts are mostly due to write-write conflicts, aborts
in Fabric come from two sources: inconsistent reads and the read-write
conflicts.
On the one hand, during the proposal phase in Fabric, a client must collect
identical read results from the peers.
This is because we mandate that each transaction proposal must be simulated and
endorsed by all peers.
But different results may be returned, as the peers have disjoint states, which
is highly likely since they commit blocks at different rates.
In this case, the client immediately aborts the transaction.
On the other hand, any of the modified records exhibiting a read-write conflict
may render the transaction invalid.
Under 10 operations per transaction, these two sources take up $14\%$ and $86\%$
of all the aborts, respectively.

\subsubsection{Effect of record size}
\label{sec:exp:txn:record_size}

\begin{figure*}[tp]
	\centering
	\begin{subfigure}{0.4\textwidth}
		\includegraphics[width=0.99\textwidth]{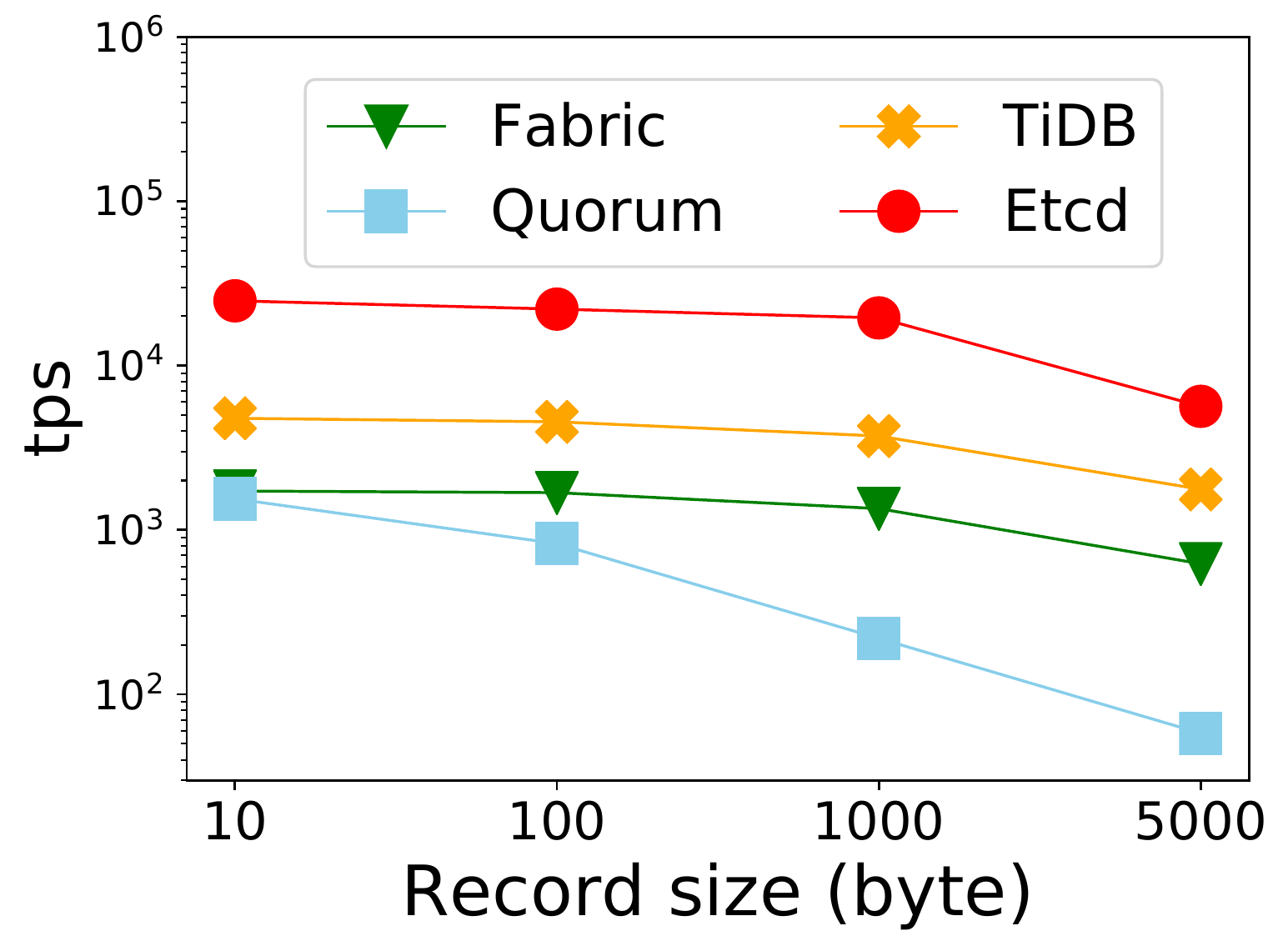}        
		\caption{Throughput}
		\label{fig:record-size-thruput}
	\end{subfigure}
	\begin{subfigure}{0.4\textwidth}
		\includegraphics[width=0.99\textwidth]{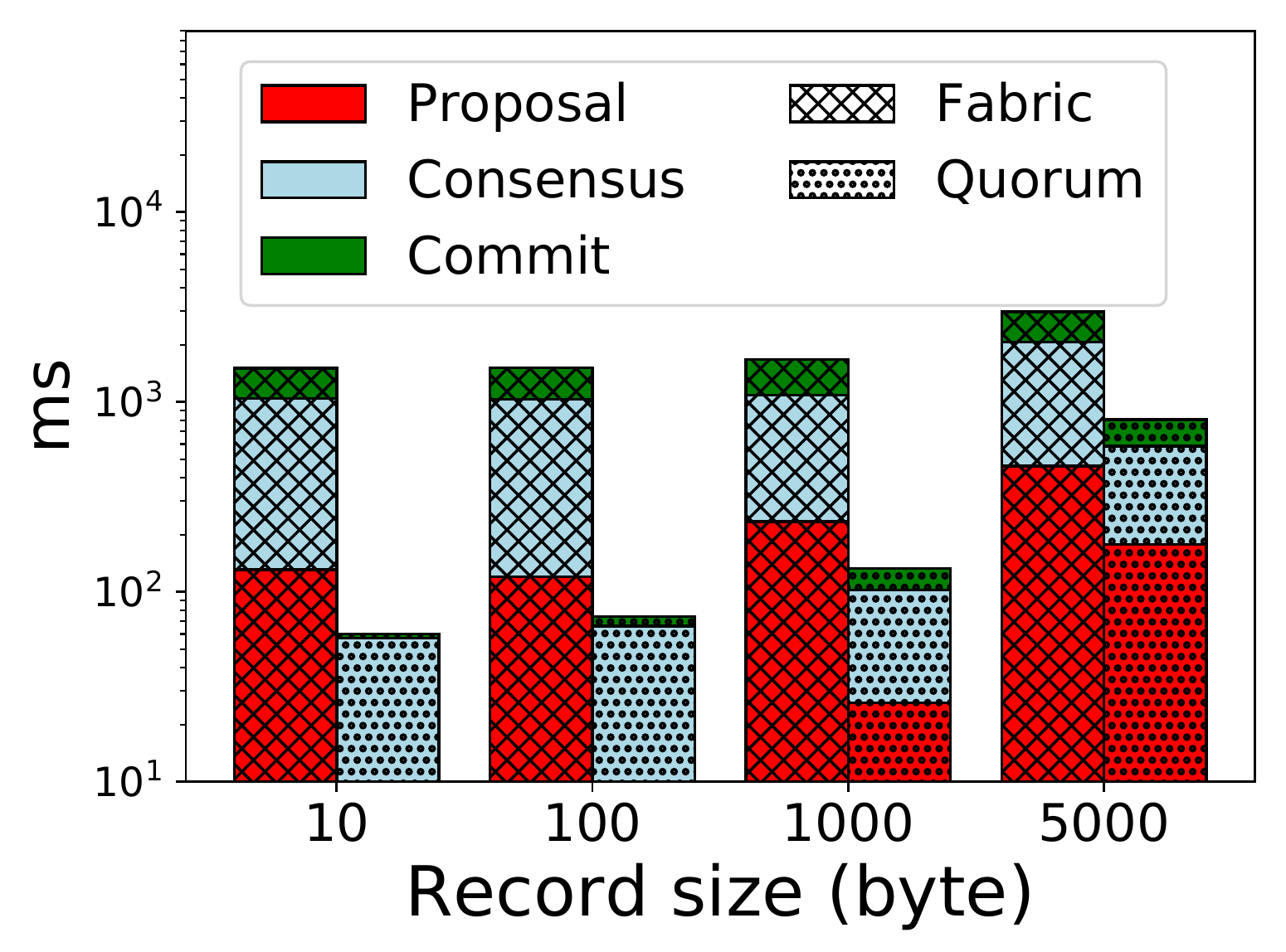}     
		\caption{Latency breakdown}
		\label{fig:record-size-breakdown}
	\end{subfigure}
	\caption{Performance under uniform update workload with increasing record size.
	Both plots use log scale.}
\end{figure*}

\begin{figure*}[tp]
	\begin{minipage}{0.325\textwidth}
		\centering
		\includegraphics[width=0.9\textwidth]{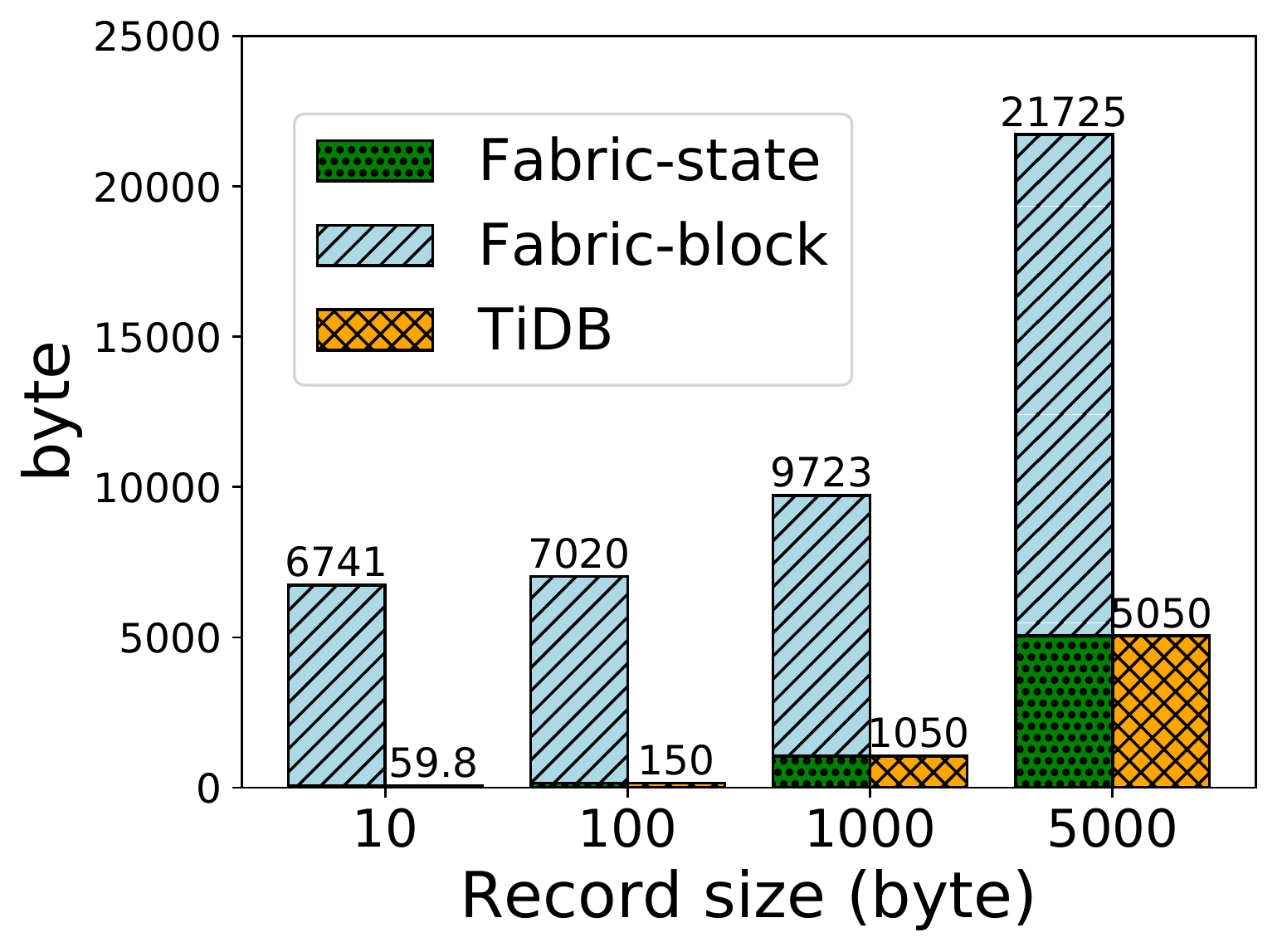}
		\caption{Storage breakdown \\ in Fabric and TiDB.}
		\label{fig:record-size-storage}
	\end{minipage}\hfill
	\begin{minipage}{0.325\textwidth}
		\centering
		\includegraphics[width=0.9\textwidth]{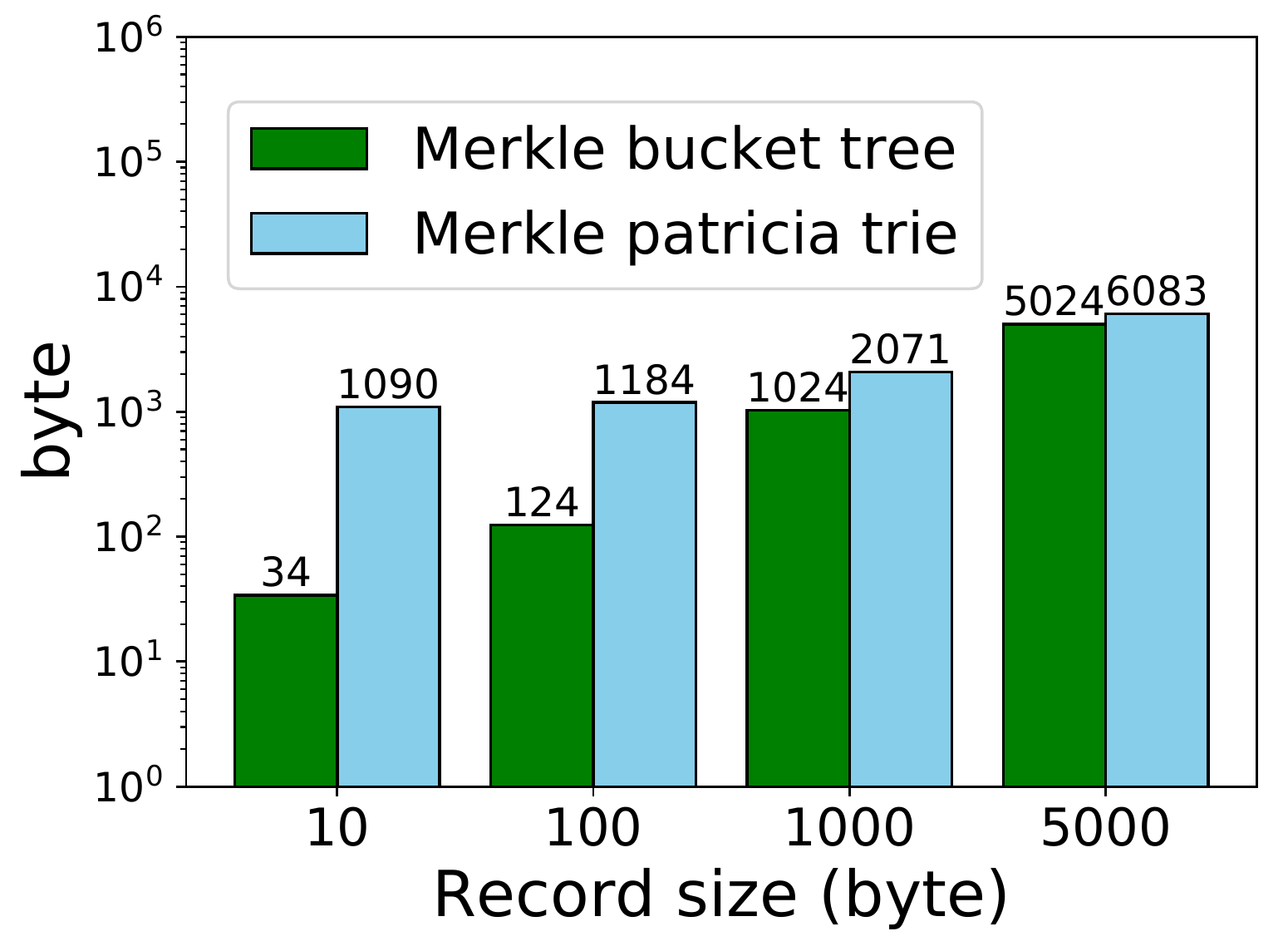}
		\caption{Storage overhead to achieve \\ tamper evidence (log scale).}
		\label{fig:record-size-antitamper}
	\end{minipage}\hfill
	\begin{minipage}{0.325\textwidth}
		\centering
		\includegraphics[width=0.9\textwidth]{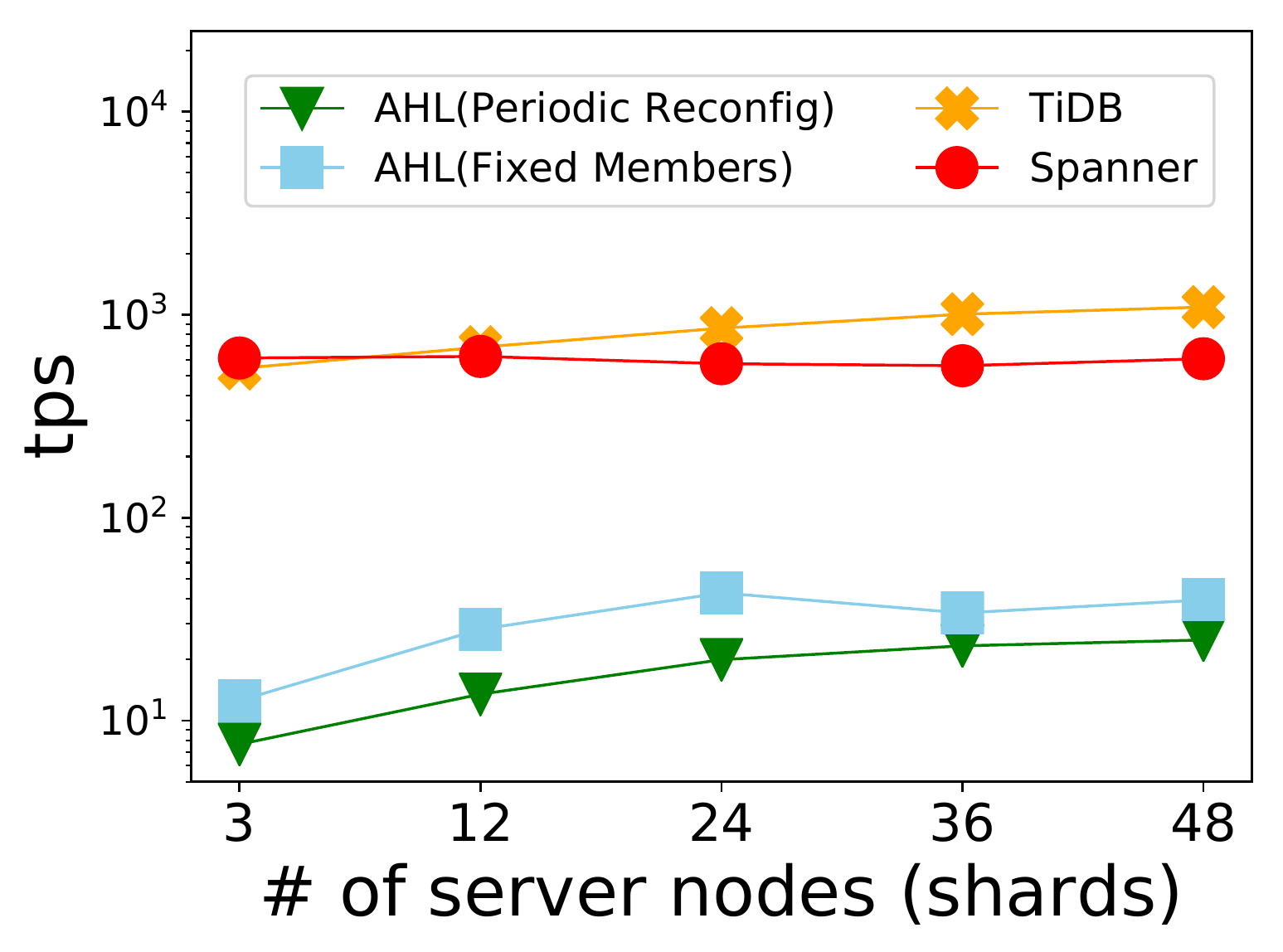}
		\caption{Throughput of the skewed workload (log scale).}
		\label{fig:shard}
	\end{minipage}
 \end{figure*}
 
We enlarge the record in the uniform-update workload to increase the complexity
per transaction without aggravating the inter-transaction conflicts.
As shown in Figure~\ref{fig:record-size-thruput}, all the databases exhibit
moderate throughput decrease and latency increase.
However, the two blockchains behave differently.
When the record grows from $10$ to $1,000$ bytes, Fabric's performance remains
roughly constant at 1400 tps and drops by half on $5,000$ bytes.
But Quorum suffers a significant drop in throughput, from $1547$ tps on
$10$-byte to $58$ tps on $5000$-byte records.
To understand this, we analyze the transaction latency breakdown in Fabric and
Quorum, and present the results in Figure~\ref{fig:record-size-breakdown}.
The block commit time in Fabric only doubles, whereas in Quorum there is a
$70\times$ increase from $3$ms for $10$-byte records to $221$ms for $5000$-byte
records, reducing the proportion of the consensus from $88\%$ to $50\%$ in a
transaction lifecycle.
For each commit, Quorum's virtual machine needs to reconstruct an MPT data
structure, which involves many expensive cryptographic hash computations.
At the same time, the cost of a hash function increases with the record size.
In particular, we find that the cost of MPT reconstruction increases from $56$us
to $2.5$ms when the record size grows from $10$ to $5000$ bytes.

Another interesting observation from Figure~\ref{fig:record-size-breakdown} is
that the delay of the proposal phase in Quorum grows at the same rate as the
delay of the commit phase.
This is due to Quorum's order-execute model, where transactions are firstly
batched and serially executed during the proposal phase by the proposer.
After consensus, the batched transactions are serially executed again by all the
other nodes for validation and commit.
Hence, Quorum's performance suffers from both double execution and the overhead
of sequential validation of in-block transactions.
In contrast, Fabric adopts an execute-order-commit model where transactions are
executed concurrently during the proposal phase, before being ordered and
batched in the consensus phase. 
The serial processing only occurs once during the commit phase. 
However, concurrency comes at the cost of potentially aborted in-block
transactions that would break the serializability, as we saw in the previous
section.
Hence, when the transactions are computationally heavy, execute-order-commit
blockchains outperform order-execute blockchains by introducing the
sequentiality requirement later.
But compared with blockchains, NewSQL databases with storage-based replication
can harness more concurrency.

\subsection{Storage}

\subsubsection{Effect of record size on storage}


Figure~\ref{fig:record-size-storage} shows the storage cost per record as we
increase the record size.
Fabric incurs a much higher storage overhead than TiDB.
For a $5000$-byte record, the state storage consumes around $5000$ bytes, while
the block storage consumes $21,725$ bytes.
There is no additional storage used by TiDB because no historical information is
maintained and the associated metadata is negligible.
This result demonstrates that blockchains incur significantly higher storage
costs than databases because of the underlying ledger abstraction.

\subsubsection{Security overhead for tamper evidence}
\label{sec:exp:storage:tamper}

To quantify the overhead incurred by the integrity protection mechanism in
blockchains, we compare the performance of Merkle Bucket Tree (MBT) from
Hyperledger Fabric v0.6\footnote{Fabric v1.0 and later relax the security model
and no longer require tamper-evident indexes.} and Merkle Patricia Trie (MPT)
from Quorum.
\revision{This comparison is done on the system behavior in its entirety. We
refer readers to~\cite{yue2020analysis} for an in-depth analysis.}

For this comparison, we insert 10K records of different sizes and measure the
state storage cost per record. 
Figure~\ref{fig:record-size-antitamper} shows
that MBT adds extra $24$ bytes per record, while MPT adds over 1KB per record.
Since both MBT and MPT store data records in the leaves, their differences come
from the tree structures: the deeper the tree, the higher the storage overhead.
The scale of MBT is fixed.
Specifically, MBT first hashes all the records into $1,000$ buckets, on top of
which a Merkle tree with a given fan-out is built.
Considering $1,000$ buckets and a fan-out of $4$ in our experiments, the depth
of the tree is capped at 5 ($\ceil{log_4 1000}$).
As a prefix tree, the depth of MPT is affected by the key length, which is 16
bytes in our setting.
Specifically, each internal MPT node holds 4 bits of the key, hence, the depth
and fan-out can go up to 32 and 16, respectively.
This explains why MPT needs more space.

\subsection{\revision{Sharding}}
\label{sec:exp:shard}

To compare the impact of sharding on databases and blockchains, we disable full
replication in TiDB, and compare its performance with Spanner, a cloud-based
NewSQL database, and Attested Hyperledger (AHL)~\cite{dang2018towards}, a
state-of-the-art sharded blockchain based on Hyperledger Fabric v0.6.
AHL leverages trusted hardware to reduce shard size and to improve throughput
per shard.
It supports cross-shard transactions by running a BFT shard that implements a
2PC state machine, and periodically reconfigures shards to mitigate adaptive
adversaries.
This experiment is run on Google Cloud Platform since Spanner is a cloud-only
service.
We set the number of nodes in a shard to 3 for all the systems, and we
pre-populate the state with 1M 1KB-size records.
We evaluate the systems with a skewed workload with a Zipfian coefficient of
$\theta=1$, in which each transaction modifies two records.

Figure~\ref{fig:shard} shows that TiDB achieves higher throughput compared to
Spanner when increasing the number of nodes (and shards).
This is because TiDB instantly aborts a transaction once detecting a conflict.
In contrast, conflicting transactions in Spanner would contend for locks under
the pessimistic concurrency control.
To achieve stronger security, AHL with periodic shard re-configuration trades
off $30\%$ in performance compared to AHL with fixed shards.
Nonetheless, the gap between AHL and both the databases is large, due to the
high cost of PBFT and other security overheads.

\subsection{\revision{Performance of Hybrid Systems}}
\label{sec:hybrids}
\begin{figure*}
	\centering
\includegraphics[width=0.79\textwidth]{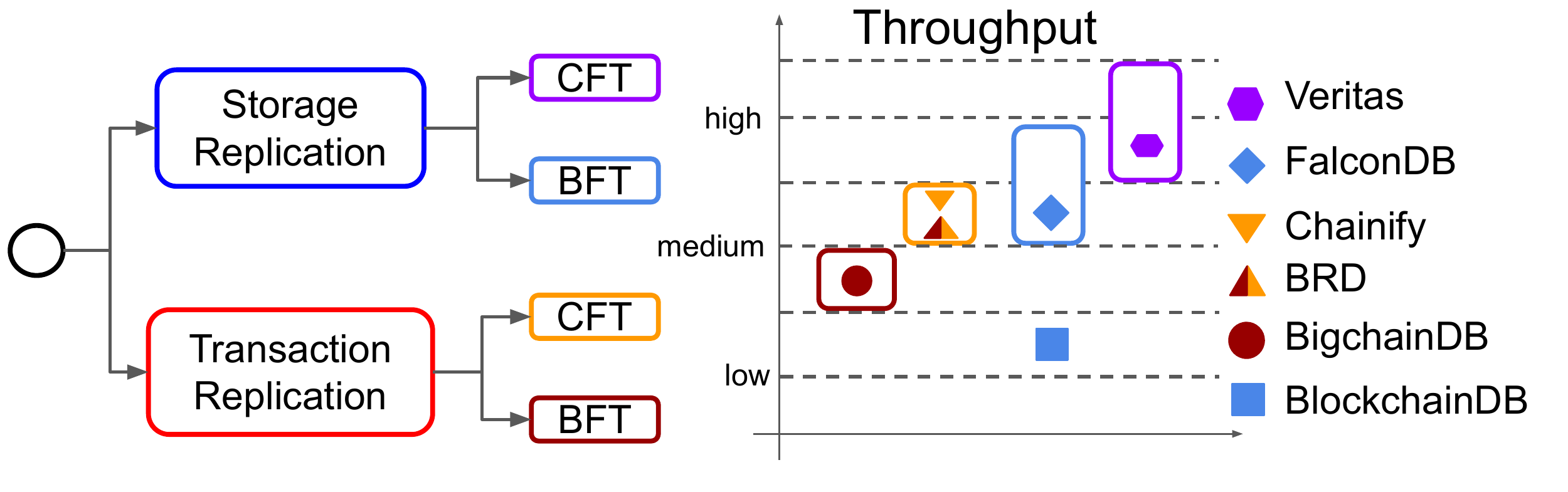}
\caption{The framework for understanding the throughput of hybrid systems.
    The systems are color-coded based on the design choices.} 
\label{fig:perf_framework}
\end{figure*}

Based on our taxonomy and experimental results, we propose a framework for
comparing the performance of existing hybrid systems.
We emphasize that the framework only supports high-level, back-of-the-envelope
comparison, and is not a replacement for detailed experimental analysis.
It focuses on throughput as the key performance metric and does not consider all
the dimensions in our taxonomy.
However, this framework explains the performance differences among systems
according to their reported results.
More importantly, it can guide the design of future hybrid systems.

Figure~\ref{fig:perf_framework} presents our framework together with the
reported performance of some hybrid systems.
We note that the replication model is the deciding factor in determining the
peak throughput.
The results in Section~\ref{sec:exp:replication:model} show that the replication
model affects concurrency.
In particular, transaction-based replication exposes lower concurrency than
storage-based replication, which results in lower throughput.
The next factor that affects throughput is the failure model.
As explained in Section~\ref{sec:taxonomy:replication:failure}, CFT protocols
are more efficient than BFT protocols due to their lower network overhead,
therefore, systems using CFT are likely to have higher throughput.
This is true especially when the CFT protocol is implemented as a shared log
service.
We note that even though our experiments do not show much difference between CFT
and BFT in Quorum, it is because these protocols are not the bottleneck.

Figure~\ref{fig:perf_framework} illustrates the reported performance of six
hybrid systems within our framework.
Using the two factors stated above, we can predict the throughput effectively.
For instance, Vertias exhibits better throughput than Chainify ($29k$ vs.
$6.1k$) because it uses storage-based replication and CFT protocols.
But its performance has a high variance because, under high contention, the
throughput can decrease significantly, as explained in
Section~\ref{sec:exp:concurrency}.

\section{Related Work}
\label{sec:related}
\textbf{Comparison.} Existing works that compare blockchains and databases have highlighted their high-level
differences. \cite{dinh2018untangling} demonstrates a significant gap in performance,
while~\cite{crosby2016blockchain,wust2018you,chowdhury2018blockchain,yaga2018blockchain}
focus on the differences at the application layer. 
Some of these studies propose empirical flow charts to guide users in the quest of choosing solutions based on blockchains or databases~\cite{wust2018you,chowdhury2018blockchain,yaga2018blockchain}.
In contrast, our work presents a deeper and more comprehensive comparison, by looking at the
fundamental designs of both systems. 

\textbf{Surveys and benchmarking.}
There are some works that conduct separate surveys and benchmarking of distributed
databases~\cite{cooper2010benchmarking,armstrong2013linkbench,abramova2013nosql,zhang2015memory} and
blockchains~\cite{thakkar2018performance,baliga2018performance,zheng2018blockchain}.  BLOCKBENCH~\cite{dinh2017blockbench} is the
first to compare them side-by-side and demonstrate that the performance of blockchain is still far behind that of distributed databases. 
Our work is more comprehensive 
than~\cite{dinh2017blockbench}, as we consider systems that are related to blockchains in their designs. We conduct more fine-grained measurements and investigate a variety of factors.

\textbf{Bridging blockchains and databases}. There is a trend of integrating database designs into blockchains and vice versa. In particular, some works apply
well-established concurrency control techniques to improve blockchain's performance \cite{dickerson2017adding, sharma2019blurring} or to reason about smart contracts' behavior~\cite{sergey2017concurrent}. \cite{wang2018forkbase,lineagechain} use database techniques
to enhance the blockchain storage layer and expose richer information to smart contracts.
\cite{BlockchainMeetsDatabase,veritas,el2019blockchaindb} propose hybrid designs that
support the relational data model and strong security. Our work provides a novel framework for exploring the design
space of hybrid, database-blockchain systems. 


\section{Conclusions}
\label{sec:conclusions}
In this paper, we presented a comprehensive dichotomy between blockchains and
distributed databases, viewing them as two different types of transactional
distributed systems. We proposed a taxonomy consisting of four design
dimensions: replication, concurrency, storage, and sharding. Using this
taxonomy, we discussed how both system types make different design choices
driven by their high-level goals, i.e., security for blockchains, and
performance for databases. We then performed a quantitative performance
comparison covering a large area of the design space.  Our results illustrate
the effects of different design choices to the overall performance.
Finally, our work provides the first framework to explore future
database-blockchain design fusions.

\section{Acknowledgements}
  The research is supported by the National Research Foundation, Singapore under its
Emerging Areas Research Projects (EARP) Funding Initiative. 
Meihui Zhang's work is supported by National Natural Science Foundation of China (62072033) and CCF- AFSG Research Fund (RF20200015). Tien Tuan Anh Dinh is supported by
Singapore University of Technology and Design's startup
grant SRG-ISTD-2019-144. 

\appendix

\bibliographystyle{abbrv}
\bibliography{main}

\end{document}